\global\def\draftcontrol{0}

   \def\versionno{ non-conformal gb}
\catcode`\@=11

\expandafter\ifx\csname draftcontrol\endcsname\relax\global\def\draftcontrol{0}
\fi

{\count255=\time\divide\count255 by 60
\xdef\hourmin{\number\count255}
\multiply\count255 by-60\advance\count255 by\time
\xdef\hourmin{\hourmin:\ifnum\count255<10 0\fi\the\count255}}
\def\draftdate{\number\month/\number\day/\number\year\ \ \ \hourmin }

\newcommand\makepapertitle{\par
  \begingroup
    \renewcommand\thefootnote{\@fnsymbol\c@footnote}%
    \def\@makefnmark{\rlap{\@textsuperscript{\normalfont\@thefnmark}}}%
    \long\def\@makefntext##1{\parindent 1em\noindent
            \hb@xt@1.8em{%
                \hss\@textsuperscript{\normalfont\@thefnmark}}##1}%
     \newpage
     \global\@topnum\z@   % Prevents figures from going at top of page.
     \@makepapertitle
     \thispagestyle{empty}\@thanks
  \endgroup
  \setcounter{footnote}{0}%
  \global\let\thanks\relax
  \global\let\makepapertitle\relax
  \global\let\@makepapertitle\relax
  \global\let\@thanks\@empty
  \global\let\@author\@empty
  \global\let\@date\@empty
  \global\let\@title\@empty
  \global\let\title\relax
  \global\let\author\relax
  \global\let\date\relax
  \global\let\and\relax
  \def\version{\let\version\@version\@gobble}
}
\def\@makepapertitle{%
  \newpage
   \ifnum\draftcontrol=1 {}
   \version\versionno
   \vskip 3em%
   \else
   \hfill\hbox to 3cm {\parbox{4cm}{\@pubnum}\hss}%
   \vskip 3em%
   \fi
   \begin{center}%
   \let \footnote \thanks
     {\LARGE {\@title}}%
     \vskip 1.5em%
     {\normalsize%\large
       \lineskip .5em%
       \begin{tabular}[t]{c}%
         \@author
       \end{tabular}\par}%
     \vskip 1.5em%
     {\@bstract}%
     \end{center}%
     \vskip 1.5em
     \@date%
   \par
}

\gdef\@pubnum{}
\def\pubnum#1{%
  \gdef\@pubnum{#1}}

\gdef\@bstract{}
\def\Abstract#1{%
  \gdef\@bstract{%
   \parbox{\textwidth-0pc}{%
   \centerline{\bf Abstract}\penalty1000%
\kern.2cm%
\noindent%\abstractfont \baselineskip=12pt
\renewcommand\baselinestretch{1.0}%
{#1}}}
}

\def\ps@paper{\let\@mkboth\@gobbletwo%
     \ifnum\draftcontrol=1
    \def\@oddfoot{\hbox to \textwidth{\tiny \versionno \hfil\tiny\draftdate}%
    \hskip -\textwidth \hbox to \textwidth{\hfil\rm\thepage\hfil}}%
     \else\def\@oddfoot{\hbox to \textwidth{\hfil\rm\thepage\hfil}}
     \fi
     \let\@evenfoot\@oddfoot
}
\def\body{\clearpage
          \pagestyle{paper}
    }
\def\@version#1{\ifnum\draftcontrol=1
\typeout{}\typeout{#1}\typeout{}
\vskip3mm\centerline{\hbox{\fbox{\normalsize{\tt DRAFT -- #1 -- }
                   {\draftdate}}}}\vskip3mm
\fi}
\let\version\@version
\long\def\eqlabel#1{\ifnum\draftcontrol=1
                    \tag@false  % there are some problems with multline without this
                    \tag*{(\theequation) \hbox to -0.2cm{\hspace{0cm}\small{#1}\hss}}
                    \refstepcounter{equation}
                    \edef\@currentlabel{\theequation}
                    \ltx@label{#1}          % use old LaTeX \label instead of new definition
                    \else
                    \label{#1}
                    \fi
                    }
\let\st@bibitem\@bibitem
\let\st@lbibitem\@lbibitem
\ifnum\draftcontrol=1
  \def\@bibitem#1{%
    \st@bibitem{#1}\a@@label{#1}\ignorespaces}
  \def\@lbibitem[#1]#2{%
    \st@lbibitem[#1]{#2}\a@@label{#2}\ignorespaces}
  \def\a@@label#1{%
    \gdef\a@lab{\smash{\normalfont\small#1}}
    \ifvmode
      \if@inlabel
        \global\setbox\@labels\hbox{%
          \llap{\a@lab\let\a@lab\relax
                \kern\@totalleftmargin\kern\marginparsep}%
          \box\@labels}%
      \fi
    \fi}
\fi
\documentclass[12pt,letterpaper]{article}
\usepackage{amsmath,amssymb,array,calc,epsfig,rotating,bm}
\usepackage[sort]{cite}
\usepackage{graphicx}
\usepackage{psfrag,verbatim}

\ifnum\draftcontrol=1
\fi
\renewcommand\baselinestretch{1.25}
\renewcommand\section{\@startsection {section}{1}{\z@}%
                                   {-3.5ex \@plus -1ex \@minus -.2ex}%
                                   {2.3ex \@plus.2ex}%
                                   {\normalfont\large\bfseries}}
\renewcommand\subsection{\@startsection{subsection}{2}{\z@}%
                                   {-3.25ex\@plus -1ex \@minus -.2ex}%
                                   {1.5ex \@plus .2ex}%
                                   {\normalfont\normalsize\bfseries}}
\renewcommand\subsubsection{\@startsection{subsubsection}{3}{\z@}%
                                   {-3.25ex\@plus -1ex \@minus -.2ex}%
                                   {1.5ex \@plus .2ex}%
                                   {\normalfont\normalsize\it}}
\renewcommand\paragraph{\@startsection{paragraph}{4}{\z@}%
                                   {-3.25ex\@plus -1ex \@minus -.2ex}%
                                   {1.5ex \@plus .2ex}%
                                   {\normalfont\normalsize\bf}}

\numberwithin{equation}{section}

\def\revise#1       {\raisebox{-0em}{\rule{3pt}{1em}}%
                     \marginpar{\raisebox{.5em}{\vrule width3pt\
                     \vrule width0pt height 0pt depth0.5em
                     \hbox to 0cm{\hspace{0cm}{%
                     \parbox[t]{4em}{\raggedright\footnotesize{#1}}}\hss}}}}

\newcommand\nxt[1]  {\\\fnxt#1}
\newcommand{\ie}{{\it i.e.,}\ }
\newcommand{\eg}{{\it e.g.,}\ }

\def\calb         {{\cal B}}

\def\cale         {{\cal E}}
\def\calf         {{\cal F}}

\def\cali         {{\cal I}}

\def\call         {{\cal L}}
\def\calm         {{\cal M}}
\def\caln         {{\cal N}}
\def\calo         {{\cal O}}
\def\calp         {{\cal P}}

\def\calt         {{\cal T}}

\def\del          {\partial}

\def\Re           {{\rm Re\hskip0.1em}}

\def\sqr#1#2{{\vcenter{\vbox{\hrule height.#2pt
 \hbox{\vrule width.#2pt height#1pt \kern#1pt
 \vrule width.#2pt}\hrule height.#2pt}}}}

\newcommand{\kk}{\mathfrak{q}}
\newcommand{\ww}{\mathfrak{w}}

\def\a{\alpha}
\def\b{\beta}
\def\w{\omega}
\def\r{\rho}
\def\dd{\delta}
\def\hdd{\hat{\delta}}
\def\e{\epsilon}

\def\hg{\hat{\gamma}}

\def\aa1{\phi}
\def\cc1{\psi}

\def\k{\kappa}

\def\l{\lambda}

\def\k{\kappa}

\def\s{\sigma}

\def\hw{\hat{\omega}}

\def\lgb{\l_{GB}}

\catcode`\@=12

\begin{document}

%%%
%%%%%% text starts here
%%%%%%%%%

\title{\bf Non-conformal holographic Gauss-Bonnet hydrodynamics }

\date{January 18, 2018}
%\date\today

\author{
Alex Buchel\\[0.4cm]
\it $ $Department of Applied Mathematics\\
\it $ $Department of Physics and Astronomy\\ 
\it University of Western Ontario\\
\it London, Ontario N6A 5B7, Canada\\
\it $ $Perimeter Institute for Theoretical Physics\\
\it Waterloo, Ontario N2J 2W9, Canada
}

\Abstract{
We study hydrodynamics of four-dimensional non-conformal holographic
plasma with non-equal central charges $c\ne a$ at the ultraviolet
fixed point. We compute equation of state, the speed of sound waves,
transport coefficients (shear and bulk viscosities), and discuss
causality.  We study the asymptotic character of the hydrodynamic
series for the homogeneous and isotropic expansion of the plasma. We
perform computations for finite $c-a\ne 0$, but to leading
nonvanishing order in the conformal symmetry breaking coupling.
}

\makepapertitle

\body

\version\versionno
\tableofcontents

\section{Introduction}\label{intro}

Holographic correspondence \cite{m1,Aharony:1999ti} provided an
opportunity to explore near- and far-from-equilibrium properties of
strongly coupled gauge theories
\cite{Son:2007vk,Chesler:2013lia} . Recently\footnote{See
\cite{Buchel:2004di,Benincasa:2005qc,Buchel:2008sh,Buchel:2008bz,Buchel:2008kd
,Kats:2007mq,Buchel:2008vz} for the early work.}, there was
been a revival of interest in holographic models of {\it conformal}
hydrodynamics, where the dual gauge theory has a finite 't Hooft
coupling, or the non-equal central charges $c-a\ne 0$
\cite{Waeber:2015oka,Grozdanov:2016vgg,Grozdanov:2016fkt,Andrade:2016rln,Grozdanov:2016zjj,
DiNunno:2017obv,Atashi:2016fai,Casalderrey-Solana:2017zyh}. 
We should emphasize that in string-theoretic top-down holographic
constructions only the leading finite 't Hooft coupling corrections
are known, and only for $\caln=4$ supersymmetric
Yang-Mills \cite{gw,Green:2003an,Paulos:2008tn}.  Since on the
gravitational side of the duality these corrections correspond to
higher-derivative corrections in the equations of motion, they can
not, even in principle, be treated beyond infinitesimally small
approximation.  On the contrary, corrections due to non-equal central
charges of a four-dimensional conformal gauge theory are encoded in
the gravitational Gauss-Bonnet (GB) coupling constant $\lgb$ (see
section \ref{model} for details), which maintains the second-order
character of the equations of motion, thus allowing for the
holographic analysis to be extended to finite values of the coupling.
Although at a technical level holographic models can be explored for finite
$\lgb$, alas, fundamentally, these models are still consistent at best
for infinitesimal values of the GB coupling \cite{Camanho:2014apa}.

The purpose of this paper is to report the results of the study of hydrodynamics of holographic
{\it non-conformal} models with a Lagrangian density $\call$. We consider holographic
renormalization group (RG) flows close to the ultraviolet (UV) fixed point,
with Lagrangian density $\call_{CFT}$ perturbed by a relevant operator
of $\calo_\Delta$ of dimension $\Delta$:
\begin{equation}
\call=\call_{CFT}+\l_{4-\Delta} \calo_\Delta\,.
\eqlabel{deform}
\end{equation}
We allow for a finite difference of the UV CFT central charges: $c-a\ne 0$.
By 'close' we mean that the local temperature $T$
of the dual gauge theory plasma is much higher than the conformal
symmetry breaking scale, \ie
\begin{equation}
\frac{|\l_{4-\Delta}|}{T^{4-\Delta}}\ll 1\,.
\eqlabel{close}
\end{equation}

The paper is organized as follows. We introduce the model in section \ref{model}.
We discuss holographic renormalization for the RG flows with $\Delta=\{2,3\}$, and present
the equilibrium equations of state of the dual plasma. In section \ref{hydro} we first compute
the shear viscosity from the retarded two-point correlation function of the equilibrium
stress-energy tensor of the boundary plasma. Next, we compute the dispersion relation of the
sound waves in plasma, and extract the speed of sound and the bulk viscosity.
In section \ref{flrw} we study the asymptotic character of the
entropy production for the homogeneous and the isotropic expansion of the plasma to all orders
in the gradient expansion parameter.
In section \ref{causality} we discuss microscopic causality of the model. 
We conclude in section \ref{summary}.

The computational frameworks are well documented in the literature
and will not be reviewed here. For reader's convenience we collect below references
to the papers adopted in the analysis:
\nxt holographic renormalization --- \cite{Buchel:2012gw};
\nxt shear viscosity from the retarded stress-energy tensor correlation functions --- 
\cite{Baier:2007ix};
\nxt sound waves in holographic plasma --- \cite{Benincasa:2005iv,Buchel:2008uu};
\nxt beyond hydrodynamics for homogeneous and isotropic expansions --- \cite{Buchel:2016cbj};
\nxt microscopic causality --- \cite{Buchel:2009tt}.

\section{Non-conformal Gauss-Bonnet holographic model}\label{model}

We define the boundary gauge theory through its dual Gauss-Bonnet
gravitational bulk model:
\begin{equation}
\begin{split}
&\cali=\frac{1}{2\ell_P^3}\int_{\calm_5} d^5 x\sqrt{-g}\left[
\call_{CFT}+\dd \call\right]\,,\\
&\call_{CFT}=\frac{12}{L^2}+R+\frac{\lgb}{2}L^2 \left(R^2-4
R_{\mu\nu}R^{\mu\nu}+R_{\mu\nu\r\s}R^{\mu\nu\r\s}\right)\,,\\
&\dd\call=-\frac 12 (\del\phi)^2 - \frac 12 m^2 \phi^2\,,
\end{split}
\eqlabel{action}
\end{equation}
where $\call_{CFT}$ is the bulk Lagrangian of the UV conformal fixed point
with central charges \cite{Nojiri:1999mh,Myers:2010jv}
\begin{equation}
\begin{split}
c&=\frac{\pi^2}{2^{3/2}}\,\frac{L^3}{\ell_P^3}\,
(1+\sqrt{1-4\lgb})^{3/2}\,\sqrt{1-4\lgb}\,,\\
a&=\frac{\pi^2}{2^{3/2}}\,\frac{L^3}{\ell_P^3}\,
(1+\sqrt{1-4\lgb})^{3/2}\,\left(3\sqrt{1-4\lgb}-2\right)\,,
\end{split}
\eqlabel{ca}
\end{equation}
and $\dd\call$ is the  conformal symmetry breaking  perturbation
realizing \eqref{deform} for
\begin{equation}
m^2 L^2 \b_2 =\Delta (\Delta-4)\,,\qquad \b_2\equiv \frac 12 +\frac 12\, \sqrt{1-4\lgb}\,,
\qquad \lgb=\b_2-\b_2^2\,.
\eqlabel{defb2}
\end{equation}
The UV conformal fixed point is a causal gauge theory, provided
\cite{Hofman:2008ar,Buchel:2009tt},
\begin{equation}
-\frac{7}{36}\le\lgb\le \frac{9}{100}\qquad
\Longleftrightarrow \qquad -\frac 12 \le \frac{c-a}{c} \le \frac 12\,.
\eqlabel{uvcausal}
\end{equation}
In what follows, without the loss of generality, we set $L=1$.

To study equilibrium thermal states of the model we use the bulk
metric ansatz
\begin{equation}
ds_5^2= \frac{r_h^2}{x}\left(-f_1 \b_2\ dt^2+\sum_{i=1}^3
dx_i^2\right)+\frac{1}{f_2} \frac{dx^2}{4x^2}\,,
\eqlabel{metric}
\end{equation}
where the metric warp factors $f_i$ and the bulk scalar $\phi$ are functions of the
radial coordinate $x$ only,
\begin{equation}
x\in (0,1)\,.
\eqlabel{defx}
\end{equation}
The asymptotically AdS boundary is located at $x=0$ and the  regular Schwarzschild horizon
at $x=1$. Parameter $r_h$ is related to the Hawking temperature of the horizon.
Asymptotically near the boundary, 
\begin{equation}
\phi=\dd_\Delta \times \begin{cases}
x^{1/2} +\calo(x^{3/2})\,,\qquad &\Delta=3\,,\\
x \ln x +\calo(x)\,,\qquad  &\Delta=2\,,
\end{cases}
\eqlabel{phib}
\end{equation}
with the non-normalizable component of the scalar field $\dd_\Delta$ identified with the corresponding
coupling constant  $\l_{4-\Delta}$ as 
\begin{equation}
\l_{4-\Delta}=\dd_\Delta r_h^{4-\Delta}\,.
\eqlabel{relatel}
\end{equation}
For a vanishing source $\dd_\Delta=0$, the above gravitational background, explicitly,
\begin{equation}
f_1=f_2\equiv f(x)=\frac{1-\sqrt{1-4(\b_2-\b_2^2)(1-x^2)}}{2(\b_2-\b_2^2)}\,,\qquad \phi=0\,,
\eqlabel{cft}
\end{equation}
describes the gravitational dual to a thermal state of the UV conformal fixed point.
For $\dd_\Delta\ne 0$ the background geometry can be easily constructed numerically. 

The equilibrium thermal state of the boundary gauge theory is characterized by the
temperature $T$, the entropy density $s$, the pressure $P$ and the energy density $\cale$. 
The entropy density is the Bekenstein or the Wald entropy\footnote{Both are the same for the GB gravity,
see \cite{Buchel:2010wf}.} of the background geometry:
\begin{equation}
s=\frac{2\pi r_h^3}{\ell_P^3}\,,
\eqlabel{s}
\end{equation}
the temperature is related to the surface gravity $\k$ at the horizon,
\begin{equation}
T=\frac{\k}{2\pi}=\frac{r_h\b_2^{1/2}}{\pi}\ \frac{\sqrt{f_1'f_2'}}{2}\bigg|_{x=1}\,.
\eqlabel{tk}
\end{equation}
To compute the energy density and the pressure, one needs to
holographically renormalize the model. This step involves
specifying the generalized Gibbons-Hawking term at the regularization
boundary $\del\calm_5$, $S_{GH}$ (see \eg \cite{Grozdanov:2016fkt}),
\begin{equation}
\begin{split}
&S_{GH}=-\frac{1}{\ell_P^3}\int_{\del\calm_5}d^4x\sqrt{-\gamma}
\left[K+ (\b_2-\b_2^2)\left(J-2 G_\gamma^{\mu\nu} K_{\mu\nu}\right)
\right]\,.
\end{split}
\eqlabel{GH}
\end{equation}
Here $\gamma_{\mu\nu}=g_{\mu\nu}-n_\mu n_\nu$ is the induced metric on the boundary,
$n^\mu$ is the unit outwards vector to the boundary and $G_\gamma^{\mu\nu}$
is the induced Einsteins tensor on the boundary. The extrinsic curvature tensor
is
\begin{equation}
K_{\mu\nu}=-\frac 12 \left(\nabla_\mu n_\nu+\nabla_\nu n_\mu\right)\,,
\eqlabel{defk}
\end{equation}
$K$ is its trace and the tensor $J_{\mu\nu}$ is defined as 
\begin{equation}
J_{\mu\nu}=\frac 13 \left(
2 K K_{\mu\r}K^\r_{\, \nu}
\right)+K_{\r\sigma}K^{\r\sigma}K_{\mu\nu}-2 K_{\mu\r}K^{\r\sigma}K_{\sigma\nu}-K^2 K_{\mu\nu}\,,
\eqlabel{defj}
\end{equation}
with $J$ being the trace of the latter. Additionally, we must include
the counter-term action at the regularization boundary (located at the radial
position $x=\e$):
\begin{equation}
\begin{split}
&S_{c.t.}=\frac{1}{\ell_P^3}\int_{\del\calm_5}d^4x\sqrt{-\gamma} \left[
\call_{c.t.,CFT}+\call_{c.t.,\Delta}\right]\,,\\
&\call_{c.t.,CFT}=-\left(2 b_2^{1/2}+b_2^{-1/2}\right)
+\left(\frac 12 b_2^{3/2}-\frac 34 \b_{2}^{1/2}\right) R_\gamma
+\left(\frac18 \b_2^{5/2}-\frac{1}{16}\b_2^{3/2}\right)\calp_{2,\gamma}\ \ln \e\,,\\
&\call_{c.t.,\Delta}=
\begin{cases}
-\frac 14\b_2^{-1/2}\phi^2-\frac{\b_2^{-1/2}}{48(2\b_2-1)}\phi^4\ \ln \e
-\frac{\b_2^{1/2}}{48} R_\gamma\phi^2\ \ln \e \,&,\qquad \Delta=3\,,\\
-\frac 12 \b_2^{-1/2}\phi^2-\frac 12\b_2^{-1/2}\phi^2\
\frac{1}{\ln\e}\,&,\qquad \Delta=2\,,
\end{cases}
\end{split}
\eqlabel{counter}
\end{equation}
where we separated the counterterms necessary to renormalize conformal
fixed point $\call_{c.t.,CFT}$, and the
deformation-dependent set of counterterms $\call_{c.t.,\Delta}$.
Here\footnote{The terms involving
the induced Ricci tensor are necessary to compute the retarded
correlation functions of the stress-energy tensor even for the
Minkowski boundary metric.}
$R^{\mu\nu}_\gamma$ is the induced Ricci tensor on the regularization
boundary and $R_\gamma$ is its trace, and
\begin{equation}
\calp_{2,\gamma}=\calp_{\gamma}^{\mu\nu} \calp_{\mu\nu,\gamma}-
\left(\gamma^{\mu\nu}\calp_{\mu\nu}\right)^2\,,\qquad \calp^{\mu\nu}_\gamma
=R_\gamma^{\mu\nu}-\frac 16 R_\gamma \gamma^{\mu\nu}\,.
\eqlabel{delp2}
\end{equation}

\begin{figure}[t]
\begin{center}
\psfrag{l}{{$\lgb$}}
\psfrag{f}{{$\calf_3$}}
\psfrag{g}{{$\calf_2$}}
  \includegraphics[width=2.4in]{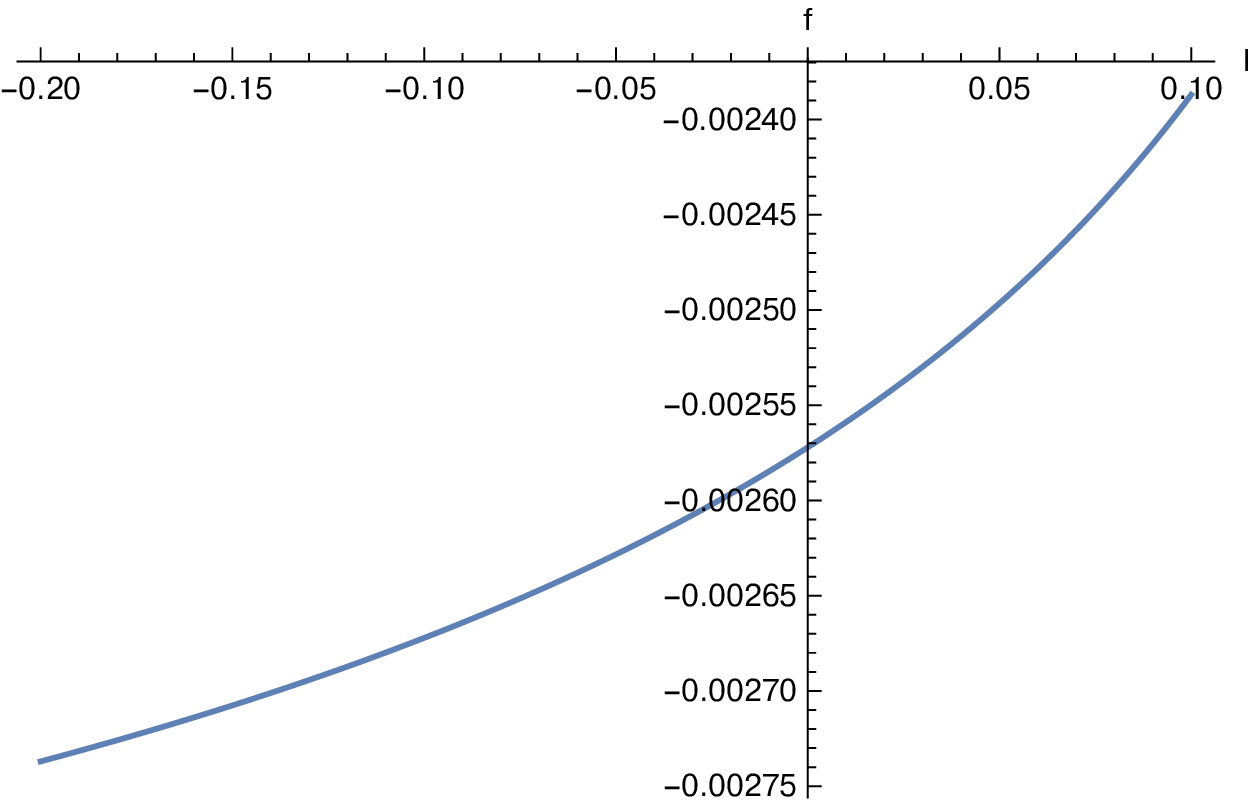}\qquad \qquad
  \includegraphics[width=2.4in]{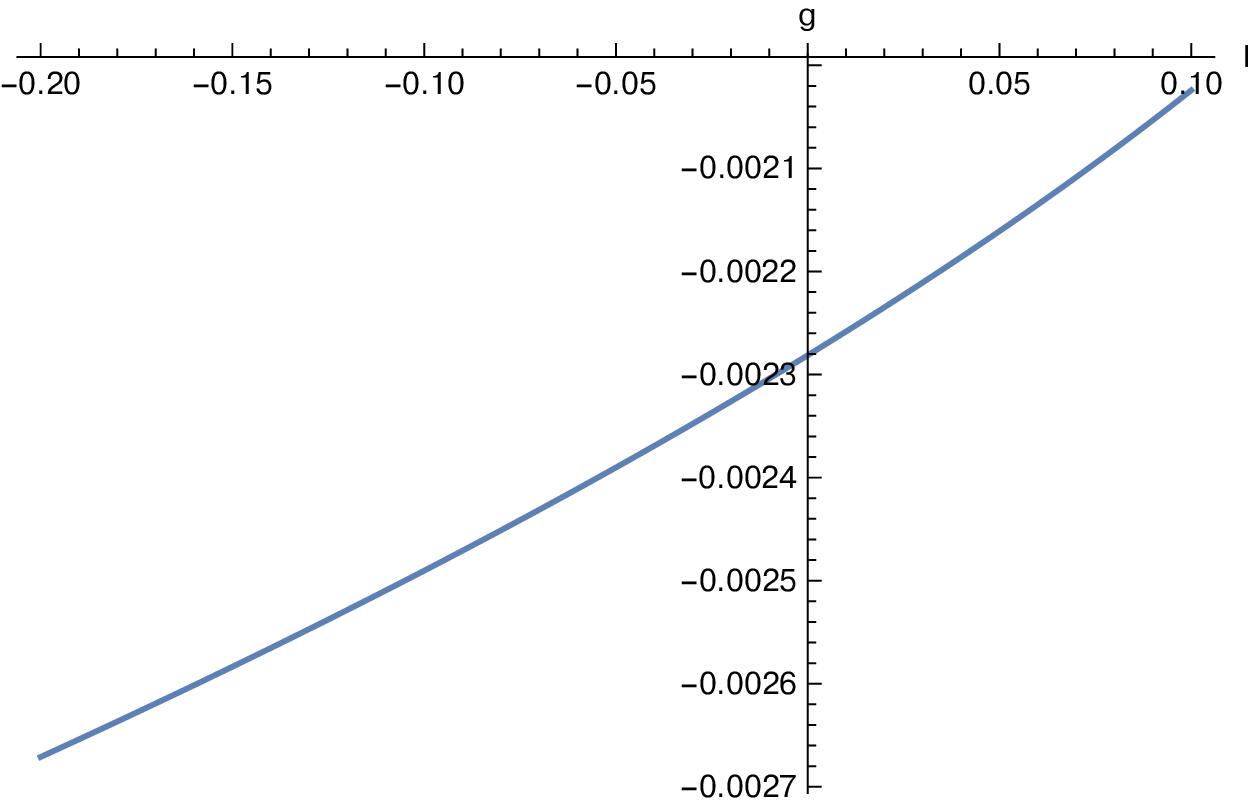}
\end{center}
 \caption{Parameterization of the equilibrium equation of state of
 the holographic non-conformal
 GB model \eqref{action} with the equation \eqref{eospar} for deformation of the
 UV conformal fixed point with $\Delta=3$ (left panel)
 and $\Delta=2$ (right panel) operators. Note that $\calf_{\Delta}\propto +(c_s^2-1/3)$ and so, 
within the causality window, the speed of sound is bounded from above by its conformal value. 
}\label{figure1}
\end{figure}

\begin{figure}[t]
\begin{center}
\psfrag{l}{{$\lgb$}}
\psfrag{f}{{$\xi_3$}}
\psfrag{g}{{$\xi_2$}}
  \includegraphics[width=2.4in]{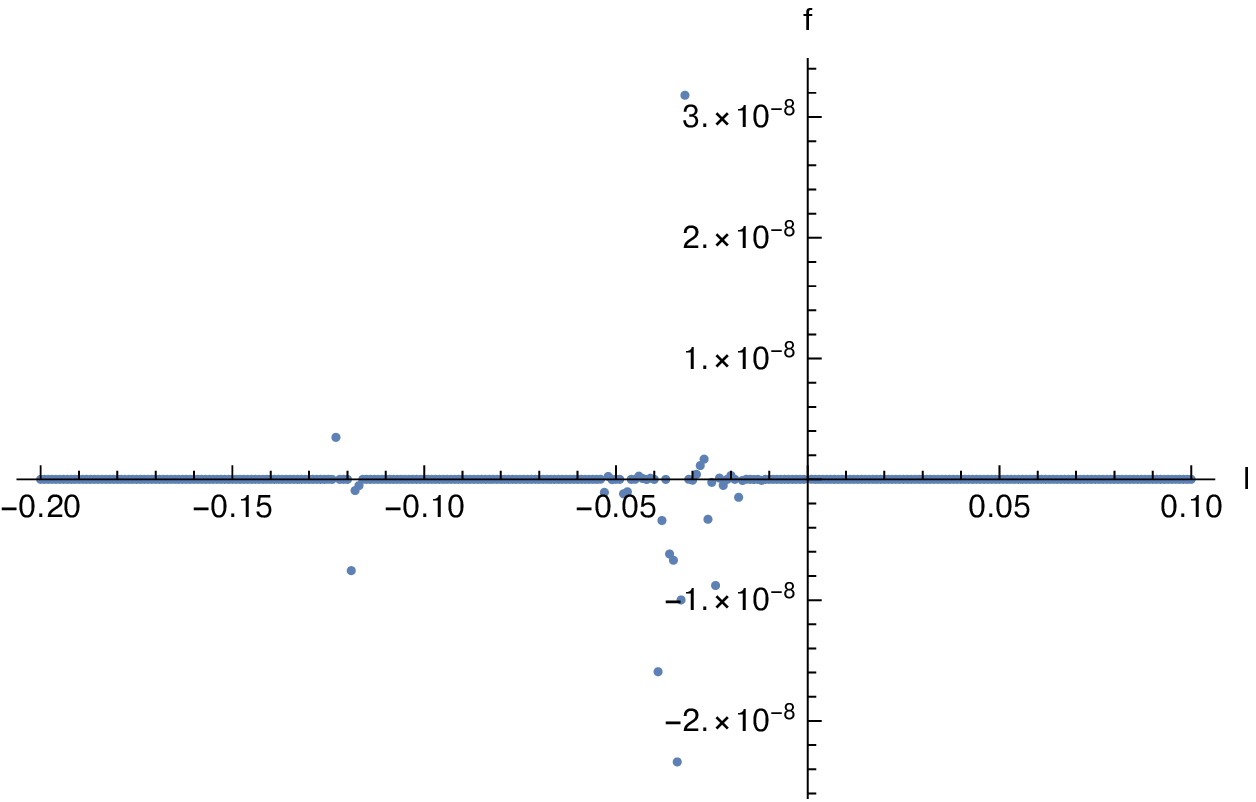}\qquad \qquad
  \includegraphics[width=2.4in]{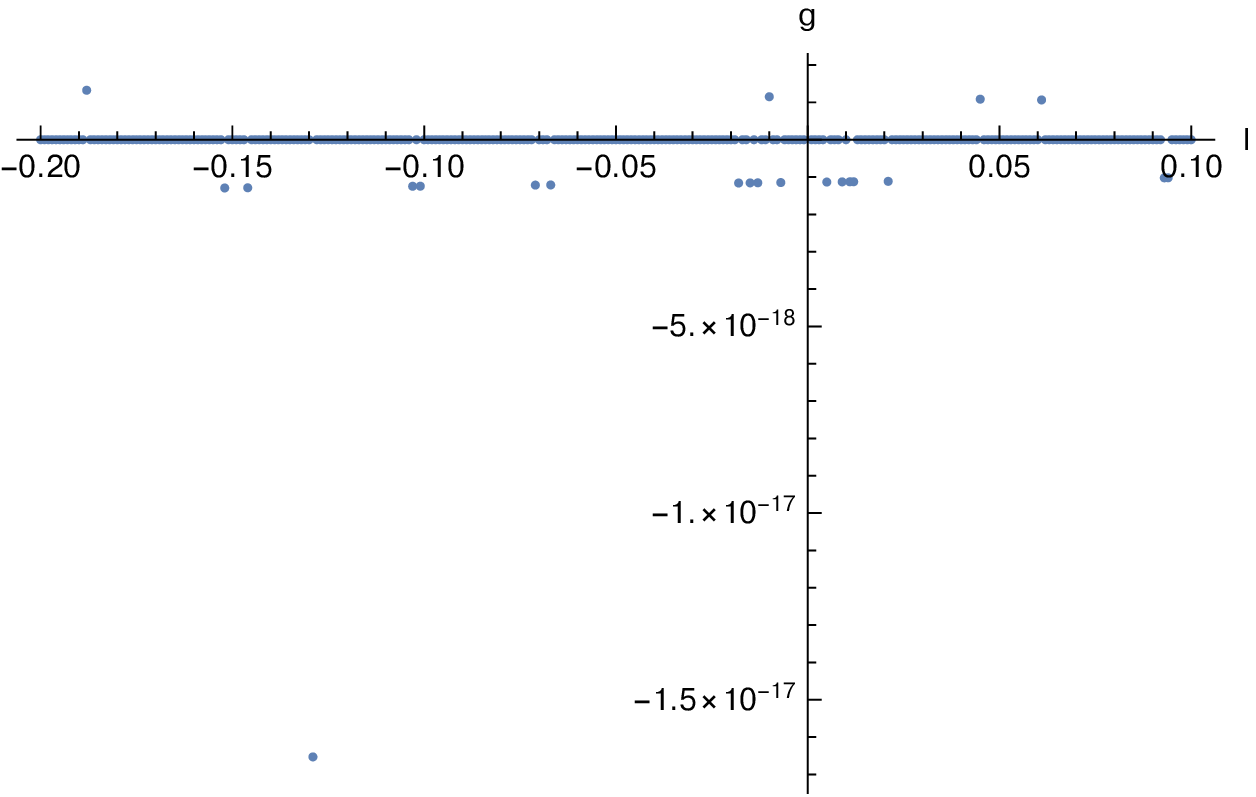}
\end{center}
 \caption{
 Numerical tests of the first law of the thermodynamics for the non-conformal GB RG  flows
 induced by $\Delta=3$ operator (left panel) and $\Delta=2$ operator (right panel).
 See \eqref{delta1} for the parameterization of the deviation $\xi_\Delta$. 
}\label{figure2}
\end{figure}

In practice, we compute the thermodynamic characteristics of the equilibrium
state to order $\calo(\dd_\Delta^2)$ inclusive. Thus, we parameterize
the non-conformal equation of state as
\begin{equation}
c_s^2-\frac 13=\left(\frac{\l_{4-\Delta}}{T^{4-\Delta}}\right)^2\ \calf_\Delta(\lgb)\,,
\eqlabel{eospar}
\end{equation}
where
\begin{equation}
c_s^2=\frac{\del P}{\del\cale}\,,
\eqlabel{defcs}
\end{equation}
is the speed of the sound waves in plasma. 
The results for $\calf_\Delta(\lgb)$ in the GB causal window \eqref{uvcausal}
are presented in fig.~\ref{figure1} for $\Delta=3$ (left panel)
and $\Delta=2$ (right panel).

While the basic thermodynamic
relation, $F$ is the free energy density, 
\[
F=-P= \cale -s T 
\]
is satisfied automatically, the first law of thermodynamics, $d\cale =T ds $, does not: it provides an
important test on our numerical data. In fig.~\ref{figure2} we present tests of the first law of the thermodynamics
\begin{equation}
\xi_\Delta(\lgb)\equiv \left(\frac{T^{4-\Delta}}{\l_{4-\Delta}}\right)^2\ \times\
\frac{1}{s}\left(\frac{d\cale}{dT}-T \frac{ds}{dT}\right)\,,
\eqlabel{delta1}
\end{equation}
for $\Delta=3$ (left panel) and $\Delta=2$ (right panel) within the GB causal window \eqref{uvcausal}.

\section{Hydrodynamic transport: shear and bulk viscosities}\label{hydro}

\begin{figure}[t]
\begin{center}
\psfrag{l}{{$\lgb$}}
\psfrag{f}{{$\eta_3$}}
\psfrag{g}{{$\eta_2$}}
  \includegraphics[width=2.4in]{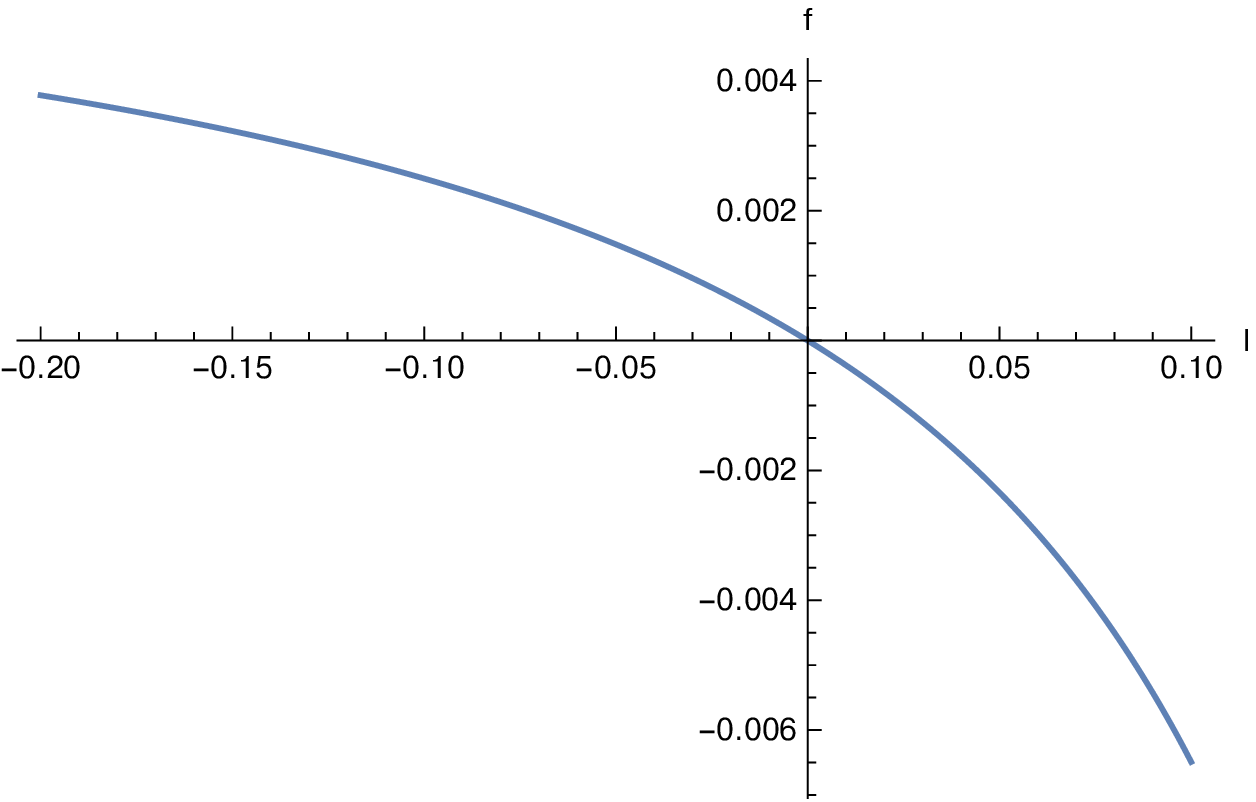}\qquad \qquad
  \includegraphics[width=2.4in]{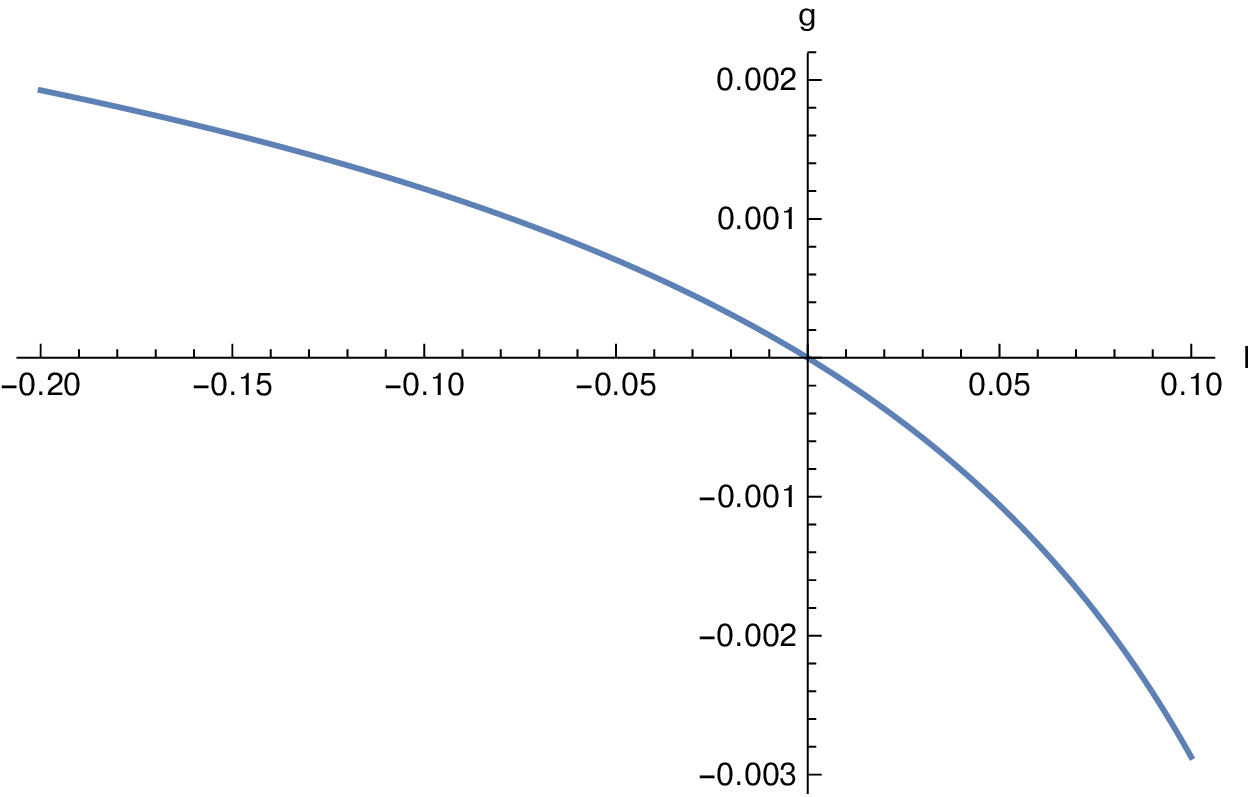}
\end{center}
 \caption{
Corrections to the shear viscosity for the non-conformal GB RG flows induced by dimension $\Delta=3$
operator (left panel) and $\Delta=2$ operator (right panel). See \eqref{etas}
for the parameterization of the corrections.
}\label{figure3}
\end{figure}

The most straightforward way to compute the shear viscosity $\eta$ of the model
is to, following \cite{Baier:2007ix}, compute the on-shell renormalized
boundary action
\begin{equation}
\begin{split}
&S_{renom}[h_{12}(t)] = \cali+S_{GB}+S_{c.t.}\bigg|_{ds_4^2}\,,\\
&ds_4^2\equiv \hg_{\a\b} dx^\a dx^\b\equiv -dt^2+\sum_{i=1}^3 dx_i^2+2h_{12}(t) dx_1dx_2\,,
\end{split}
\end{equation}
(see \eqref{action}, \eqref{GH} and \eqref{counter})
to quadratic order in the boundary metric source term $h_{12}(t)$.
The thermal expectation value of the boundary stress-energy tensor
\begin{equation}
\langle T^{\a\b}\rangle\bigg|_T = \frac{2}{\sqrt{-\hg}}\ \frac{\dd S_{renom}}{\dd \hg_{\a\b}}\,,
\eqlabel{bset}
\end{equation}
in the low frequency limit, \ie
\begin{equation}
T \bigg|\frac {\dot{h}_{12}}{h_{12}}\bigg|\ll 1\,,
\eqlabel{lowfreq}
\end{equation}
then has the off-diagonal component
\begin{equation}
T^{12}=\biggl(-P h_{12}-\eta \dot{h}_{12}+\calo\left(\ddot{h}_{12}\right)\biggr)
+\calo\left(h_{12}^2\right)\,,
\eqlabel{toff}
\end{equation}
allowing for the  extraction of  the shear viscosity. Results of this
tedious computation, to order $\calo(\dd_\Delta^2)$, in the parameterization 
\begin{equation}
\frac{\eta}{s}=\frac{(2\b_2-1)^2}{4\pi}\ \left(1+ \eta_\Delta(\lgb)\
\left(\frac{\l_{4-\Delta}}{T^{4-\Delta}}\right)^2\right)\,,
\eqlabel{etas}
\end{equation}
are presented in fig.~\ref{figure3} for non-conformal RG flow with $\Delta=3$
(left panel) and $\Delta=2$ (right panel). Note that $\l_{4-\Delta}=0$
result reproduced computations of  \cite{Kats:2007mq}; furthermore,
$\eta_\Delta(\lgb=0)=0$, reflecting the universality of the shear viscosity
\cite{Buchel:2003tz}.

\begin{figure}[t]
\begin{center}
\psfrag{l}{{$\lgb$}}
\psfrag{f}{{$\dd\calf_3$}}
\psfrag{g}{{$\dd\calf_2$}}
  \includegraphics[width=2.4in]{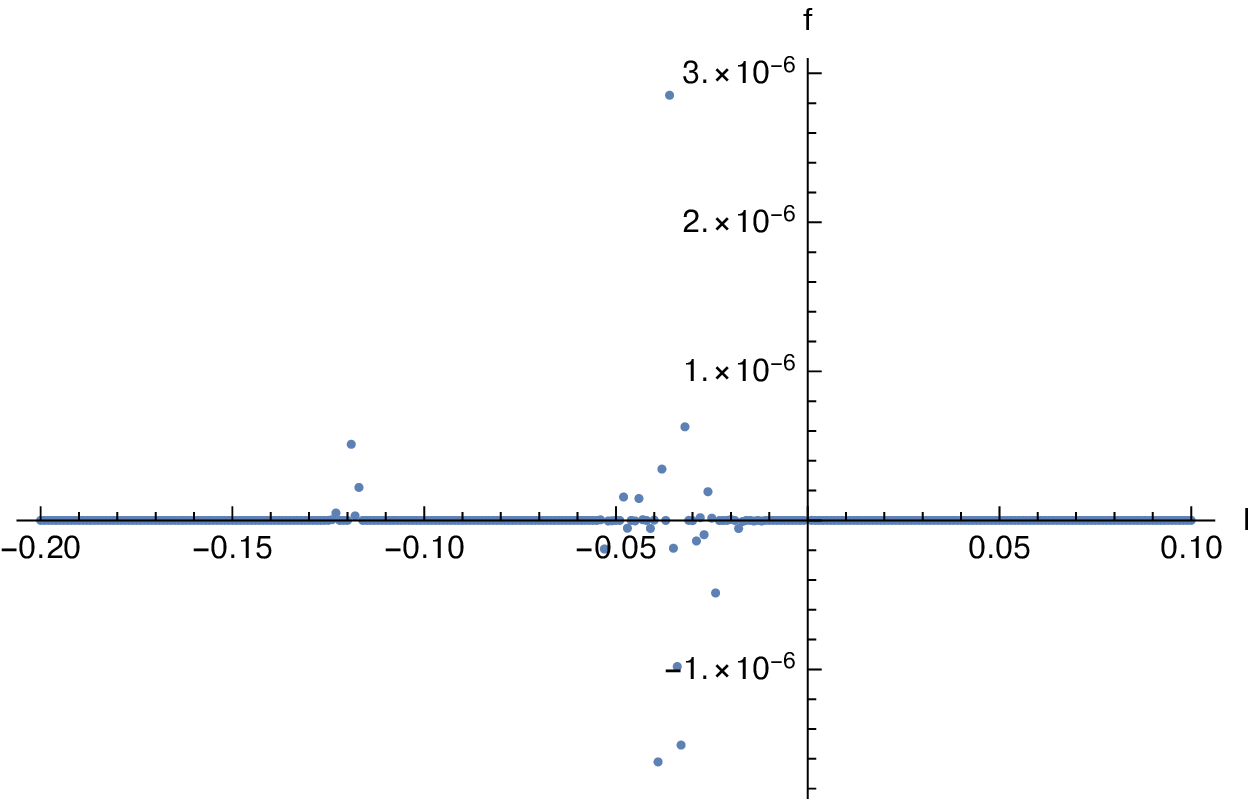}\qquad \qquad
  \includegraphics[width=2.4in]{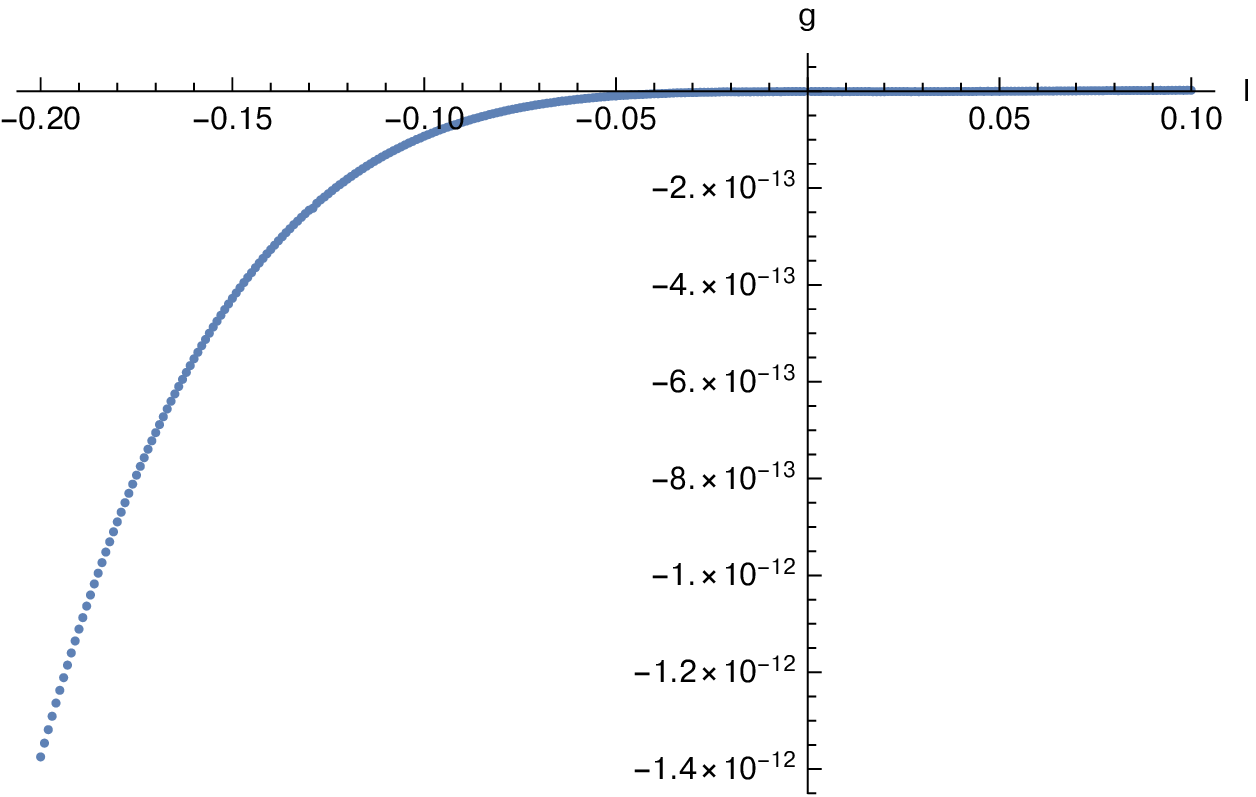}
\end{center}
 \caption{
Consistency test on extracting speed of the sound waves from the equation of state \eqref{defcs}
and directly from the dispersion relation \eqref{sounddisp}, see \eqref{deltaf}, for $\Delta=3$ RG flow
(left panel) and $\Delta=2$ RG flow (right panel). 
}\label{figure4}
\end{figure}

\begin{figure}[t]
\begin{center}
\psfrag{l}{{$\lgb$}}
\psfrag{f}{{$\zeta_3$}}
\psfrag{g}{{$\zeta_2$}}
  \includegraphics[width=2.4in]{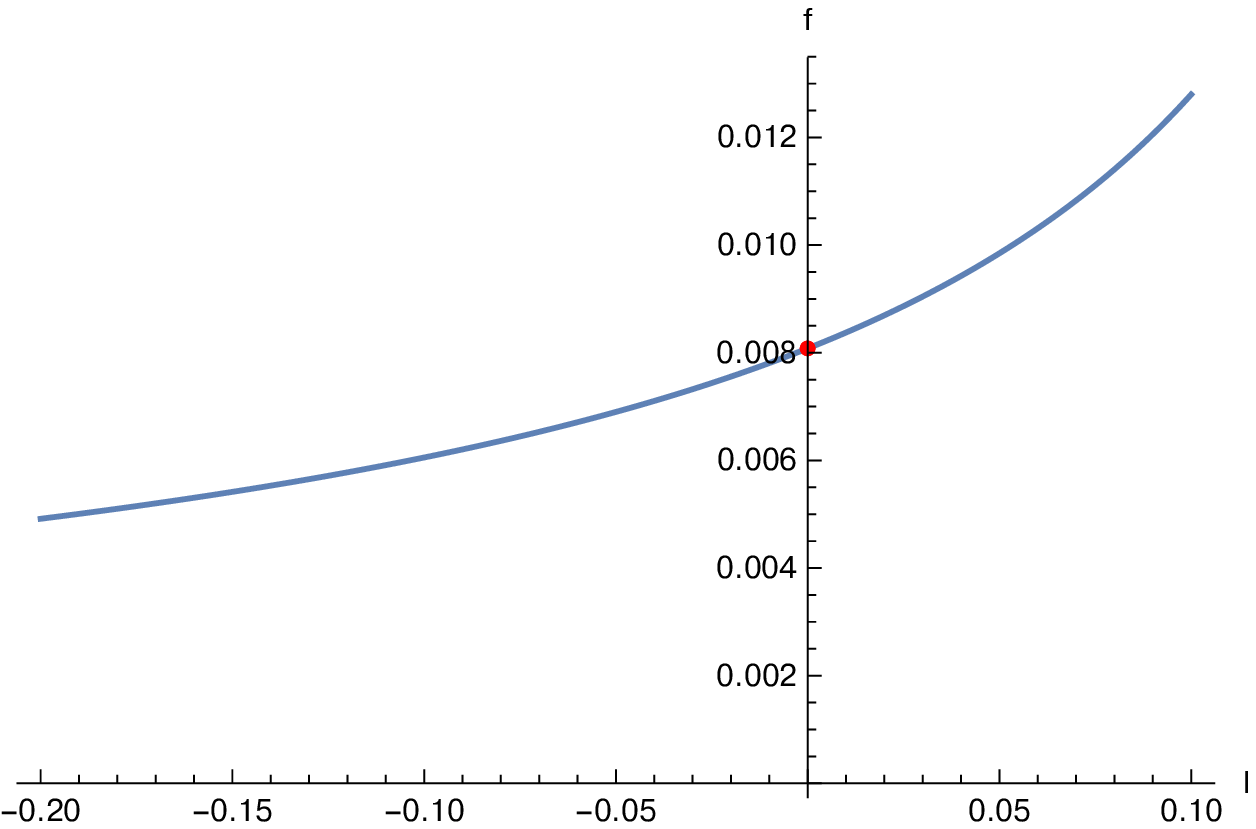}\qquad \qquad
  \includegraphics[width=2.4in]{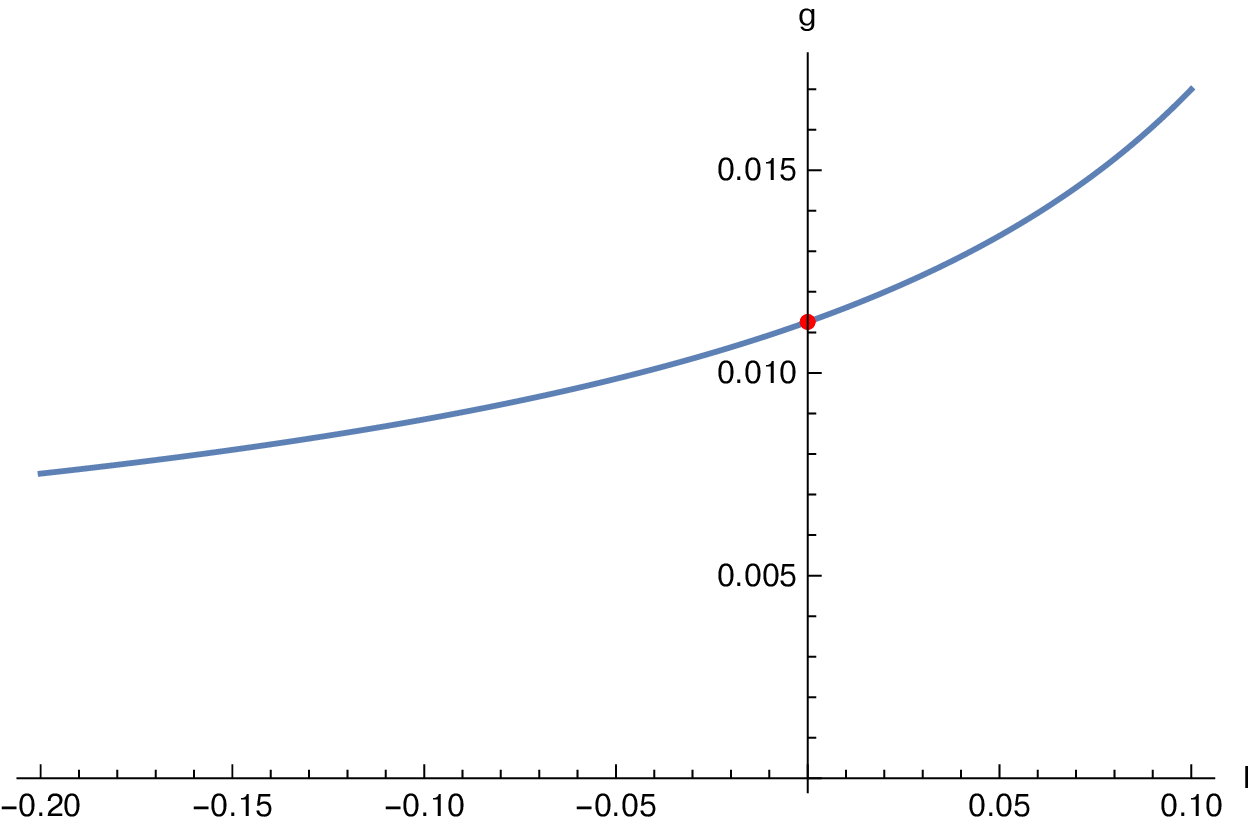}
\end{center}
 \caption{ Bulk viscosity of the holographic RG flows induced by dimension $\Delta=3$ (left panel)
 and $\Delta=2$ (right panel) operators, see \eqref{cszeta}. The red dots indicate the
 $\lgb=0$ results obtained in \cite{Benincasa:2005iv,Buchel:2008uu}.  
}\label{figure5}
\end{figure}

\begin{figure}[t]
\begin{center}
\psfrag{l}{{$\lgb$}}
\psfrag{f}{{$\calb_3$}}
\psfrag{g}{{$\calb_2$}}
  \includegraphics[width=2.4in]{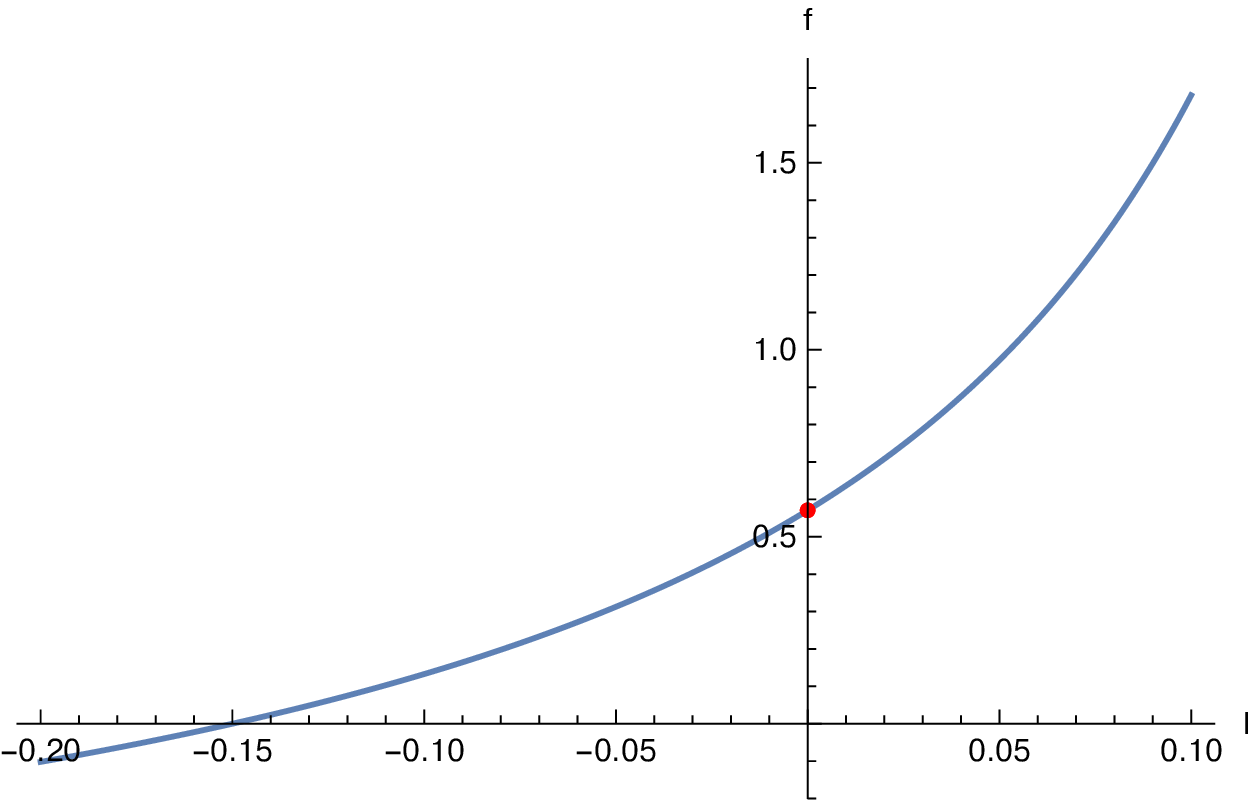}\qquad \qquad
  \includegraphics[width=2.4in]{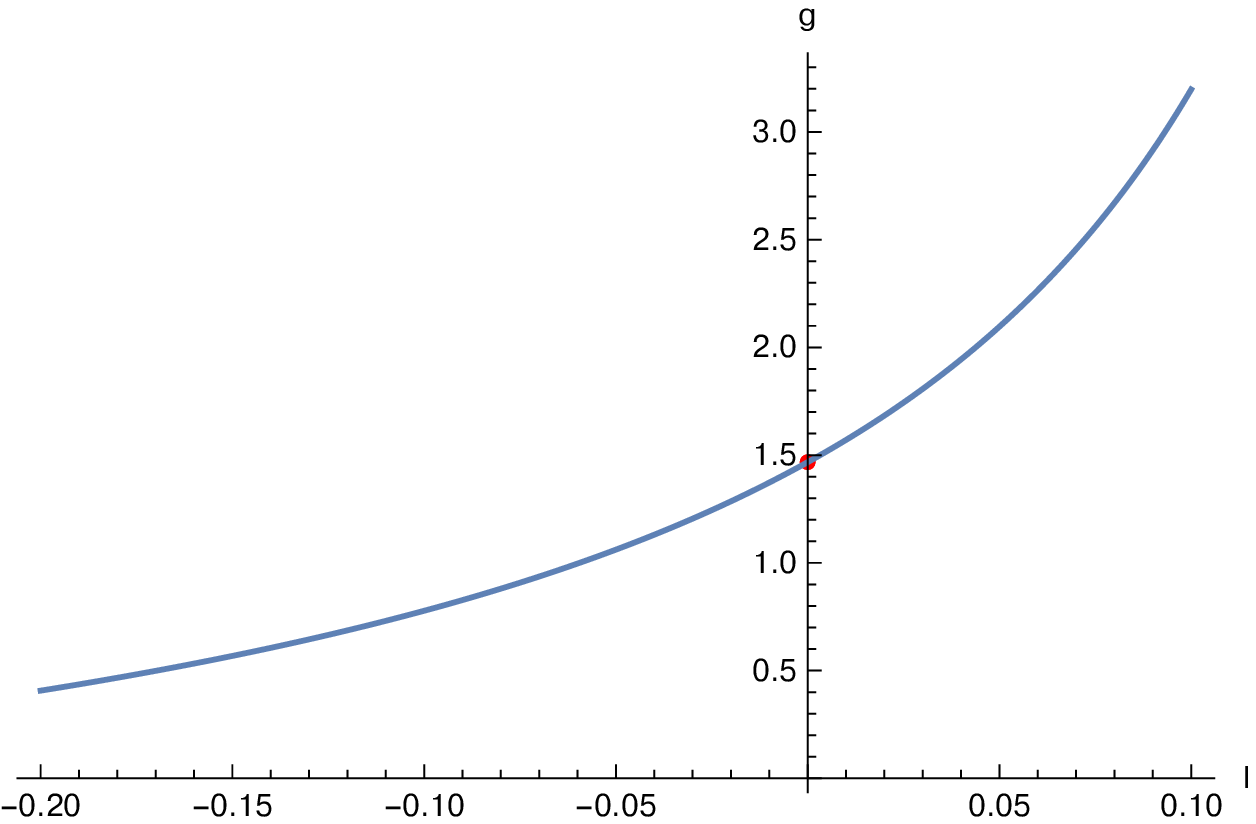}
\end{center}
 \caption{ Holographic bulk viscosity bound for RG flows induced by dimension $\Delta=3$
 (left panel) and $\Delta=2$ (right panel) operators. The bound is violated whenever $\calb_\Delta<0$.
 Red dots represent the bound at $\lgb=0$, see \eqref{bound0}. 
}\label{figure6}
\end{figure}

The spectrum of long-wavelength fluctuations in near equilibrium plasma includes
longitudinal (sound) waves with the dispersion relation:
\begin{equation}
\begin{split}
&\ww=\pm c_s\ \kk- 2\pi i\ \frac{\eta}{s}\ \left(\frac 23+\frac{\zeta}{2\eta}\right)\ \kk^2+\calo\left(\kk^3\right)\,,\\
&\ww\equiv \frac{\w}{2\pi T}\,,\qquad \kk=\frac{|\vec{q}|}{2\pi T}\,.
\end{split}
\eqlabel{sounddisp}
\end{equation}
We proceed computing the spectrum of sound waves in model \eqref{action} following \cite{Benincasa:2005iv}.
To this end we parameterize the transport coefficients as 
\begin{equation}
\begin{split}
c_s^2=\frac 13 +\left(\frac{\l_{4-\Delta}}{T^{4-\Delta}}\right)^2\ \hat{\calf}_{\Delta}(\lgb)\,,\qquad
\frac{\zeta}{\eta}=\left(\frac{\l_{4-\Delta}}{T^{4-\Delta}}\right)^2\ {\zeta_\Delta}(\lgb)\,.
\end{split}
\eqlabel{cszeta}
\end{equation}
Of course, consistency of the hydrodynamics requires that
\begin{equation}
\hat{\calf}_{\Delta}(\lgb)={\calf}_{\Delta}(\lgb)\,,
\eqlabel{consi}
\end{equation}
where the latter is introduced from the equilibrium equation of state of the plasma, following \eqref{eospar} and
\eqref{defcs}. Fig.~\ref{figure4} presents
\begin{equation}
\dd\calf_\Delta (\lgb)\equiv \frac{\hat{\calf}_{\Delta}}{{\calf}_{\Delta}} -1
\eqlabel{deltaf}
\end{equation}
 --- an important consistency check on our numerical results for RG flows with $\Delta=3$ (left panel) nd $\Delta=2$
(right panel).

To extract the bulk viscosity coefficients $\zeta_\Delta$, following \eqref{cszeta},
one has to use the results for the shear viscosity, see \eqref{etas}. Fig.~\ref{figure5}
presents  $\zeta_\Delta$ for RG flow with $\Delta=3$ (left panel) and $\Delta=2$ (right panel).
The red dots represent $\lgb=0$ results obtained in \cite{Benincasa:2005iv,Buchel:2008uu}.

We conclude this section commenting on the holographic bulk viscosity bound \cite{Buchel:2007mf},
\begin{equation}
\frac{\frac{\zeta}{\eta}}{2 \left(\frac13-c_s^2\right)}-1\ \equiv\  \calb_\Delta(\lgb)\ \ge 0 \,.
\eqlabel{bound}
\end{equation}
Note that  \cite{Buchel:2007mf}
\begin{equation}
\calb_{\Delta}\bigg|_{\lgb=0}=\begin{cases}
\frac\pi 2-1\,,\qquad &\Delta=3
\,,\\
\frac{\pi^2}{4}-1\,,\qquad  &\Delta=2\,.
\end{cases}
\eqlabel{bound0}
\end{equation}
Results for $\calb_\Delta$ are presented in fig.~\ref{figure6} for holographic RG flows with $\Delta=3$
(left panel) and $\Delta=2$ (right panel). Note that the bound is violated, within the causal window of the
model \eqref{uvcausal} for the RG flow induced by $\Delta=3$ operator (but not in the $\Delta=2$ case).
The violation happens in the theories with the UV fixed point with $a-c>0$ central charges\footnote{ The
shear viscosity bound \cite{Kovtun:2004de} is violated for CFTs with $c-a>0$ \cite{Kats:2007mq}.}. 
This is not the first known violation of the bulk viscosity bound: see \cite{Buchel:2011uj}
for the violation of the bound in a top-down model of the gauge/gravity correspondence.

\section{Homogeneous and isotropic expansion of the plasma}\label{flrw}

We study in this section homogeneous and isotropic expansion of the non-conformal plasma
defined via the dual gravitational action \eqref{action}. We follow discussion\footnote{Some related 
work appeared in \cite{Camilo:2016kxq}.} in
\cite{Buchel:2016cbj}. The purpose of the analysis is twofold:
\nxt we would like to have an independent computation of the bulk viscosity;
\nxt we would like to understand the interplay between the large-order behavior of the hydrodynamic
expansion and causality. 

Homogeneous and isotropic expansion of the boundary gauge theory plasma can be studied placing the theory in
Friedmann-Lemaitre-Robertson-Walker (FLRW) Universe with zero spatial curvature:
\begin{equation}
ds_4^2=\hg_{\a\b} dx^\a dx^\b=-dt^2 +a(t)^2\ \sum_{i=1}^3 dx_i^2\,.
\eqlabel{flrwm}
\end{equation}
In the FLRW geometry the matter expansion is locally static $u^\a=(1,0,0,0)$ though it possesses a
nonzero expansion rate $\Theta\equiv \nabla_\a u^\a=3\dot a/a$. 
The corresponding gravitational geometry is best to analyze in infalling Eddington-Finkelstein
coordinates:
\begin{equation}
ds_5^2=2dt\ \left(dr-Adt\right)+\Sigma^2\ \sum_{i=1}^3 dx_i^2\,.
\eqlabel{flrwgeom}
\end{equation}
Here, the bulk scalar field $\phi$ and the metric  warp factors $A,\Sigma$ depend only on $\{r,t\}$. 
The near-boundary $r\to \infty$ asymptotic behaviour of the metric and the scalar encode the
boundary metric scale factor $a(t)$ and the coupling constant $\l_{4-\Delta}$,
see\footnote{We use the same normalization of the couplings on the gravitational side to insure appropriate
comparison of the bulk viscosities.} \eqref{deform},
\begin{equation}
\begin{split}
&\Sigma=\frac{a}{r}+\calo(r^{-1})\,,\qquad A=\frac{r^2}{2\b_2}-\frac{\dot a r}{a }+\calo(r^0)\,,\\
&\phi=\l_{4-\Delta}
\begin{cases}
\frac 1r+\calo\left(r^{-2}\right)\,,\qquad &\Delta=3\,,\cr
-\frac{\ln r^2}{r^2} +\calo\left(r^{-2}\right)\,,\qquad &\Delta=2 \,.
\end{cases}
\end{split}
\eqlabel{boundaryfrw}
\end{equation}
As in \cite{Buchel:2016cbj} we identify the non-equilibrium entropy density $s$ with the
Bekenstein-Hawking entropy of the apparent horizon in the geometry \eqref{flrwgeom},
\begin{equation}
a^3 s=\frac{2\pi}{\ell_P^3}\ \Sigma^3\bigg|_{r=r_h}\,,
\eqlabel{eah}
\end{equation}
where $r_h$ is the location of the apparent horizon determined  from 
$d_+\Sigma|_{r=r_h}=0$ with $d_+\equiv \del_t+A\del_r$, see \cite{Chesler:2013lia}.
Taking the time derivative of the entropy density and using holographic equations of motion we find
\begin{equation}
\frac{d(a^3 s)}{dt}=\frac{4\pi}{\ell_P^3}\ (\Sigma^3)' \  
\frac{ (d_+\phi)^2}{24-m^2\phi^2}\bigg|_{r=r_h}\,.
\eqlabel{rate}
\end{equation}
Following \cite{Buchel:2017pto} it is easy to show that the gravitational equations of motion
guarantee that the entropy production rate is
nonnegative. In the hydrodynamics regime,  \ie for small Knudsen numbers $K_N =\Theta/T \ll 1$,
\begin{equation}
\frac{d}{dt}\ln \left(a^3 s\right)\bigg|_{hydro} \approx \frac 1T \left(\nabla\cdot u\right)^2\ \frac{\zeta}{s}\,,
\eqlabel{bulkflrw}
\end{equation}
provides an independent computation of the bulk viscosity.

Notice that the entropy production rate is quadratic in the bulk scalar field,
so the latter can be used in the probe approximation. Neglecting the scalar field backreaction,
\begin{equation}
\begin{split}
A&=-\frac{\dot a}{x a}+\frac{1}{4x^2 \b_2(1-\b_2)}\biggl(1-\sqrt{(2\b_2-1)^2-\frac{4x^4\b_2(\b_2-1)\mu^4}{a^4}}\biggr)\,,\\
\Sigma&=\frac{a}{x}\,,
\end{split}
\eqlabel{nonbackreaction}
\end{equation}
where we set $x\equiv \frac 1r$.
The constant parameter $\mu$ is related to the local temperature $T=T(t)=\frac{\mu}{\pi a(t)}$, and the apparent horizon is
located at
\begin{equation}
r_h=\frac{\mu}{a(t)}\,.
\eqlabel{rh}
\end{equation}
Given \eqref{nonbackreaction}, the equation of motion for the scalar field
\begin{equation}
\phi=\phi\left(t,z\equiv \frac{\mu x}{a}\right)\,,\qquad z\in (0,1)\,,
\eqlabel{defz}
\end{equation}
takes the form
\begin{equation}
\begin{split}
0=&\frac{\del^2\phi}{\del z^2}+\frac{4 a \b_2 (\b_2-1)}{\mu (1-\sqrt{G })}\
\frac{\del^2\phi}{\del t\del z}+\frac{(\sqrt{G} (3-\sqrt{G})-2 (2 \b_2-1)^2)}{z (\sqrt{G}-1) \sqrt{G}}
\ \frac{\del\phi}{\del z}\\
&+\frac{6 \b_2 a (\b_2-1)}{z \mu (\sqrt{G}-1)}\ \frac{\del\phi}{\del t} -
\frac{2 \Delta (\Delta-4) (\b_2-1)}{(\sqrt{G}-1) z^2}\ \phi\,,
\end{split}
\eqlabel{scalarflrw}
\end{equation}
where 
\begin{equation}
G\equiv (2\b_2-1)^2-4 z^4 \b_2(\b_2-1)\,.
\eqlabel{defg}
\end{equation}
A general solution of \eqref{scalarflrw} can be represented as a series expansion in the
successive derivatives of the FLRW boundary metric scalar factor $a(t)$:
\begin{equation}
\phi_\Delta=\hdd_\Delta\ a^{4-\Delta}\sum_{n=0}^\infty
\ \frac{\calt_{\Delta,n}[a]}{\mu^n}\ F_{\Delta,n}(z)\,,\qquad \hdd\equiv\frac{\l_{4-\Delta}}{\mu^{4-\Delta}}\,,
\eqlabel{series}
\end{equation}
with $\calt_{\Delta,0}=1$ and 
\begin{equation}
\calt_{\Delta,n}=\frac 14\biggl(a \dot{\calt}_{\Delta,n-1}+(4-\Delta)\dot a \calt_{\Delta,n-1}\biggr)\,,\qquad n\ge 1\,,
\eqlabel{tn}
\end{equation}
and
\begin{equation}
\begin{split}
&0=F_{\Delta,0}''+\frac{\sqrt{G} (3-\sqrt{G})-2 (2 \b_2-1)^2}{z (\sqrt{G}-1) \sqrt{G}}\
F_{\Delta,0}'-\frac{2 \Delta (\Delta-4) (\b_2-1)}{(\sqrt{G}-1) z^2}\ F_{\Delta,0}\,,\\
&0=F_{\Delta,n}''+\frac{\sqrt{G} (3-\sqrt{G})-2 (2 \b_2-1)^2}{z (\sqrt{G}-1) \sqrt{G}}\
F_{\Delta,n}'-\frac{2 \Delta (\Delta-4) (\b_2-1)}{(\sqrt{G}-1) z^2}\ F_{\Delta,n}\\
&-\frac{16 \b_2 (\b_2-1)}{\sqrt{G}-1} \left(F_{\Delta,n-1}'-\frac{3}{2z} F_{\Delta,n-1}\right)\,,\qquad n\ge 1\,,
\end{split}
\eqlabel{feqs}
\end{equation}
with boundary conditions 
\begin{equation}
F_{\Delta,0}=\begin{cases}
z+\calo(z^2)\,,\qquad &\Delta=3\,,\\
z^2\ln z^2 +\calo(z^2)\,,\qquad &\Delta=2\,,
\end{cases}\qquad F_{\Delta,n\ge 1}=\calo\left(z F_{\Delta,0}\right)\,.
\eqlabel{fbc}
\end{equation}
Recursive equations \eqref{tn} can be solved analytically for simple boundary
cosmological models \cite{Buchel:2016cbj}. Here, we will be concerned with the
de-Sitter expansion at the boundary, \ie $a(t)=e^{H t}$ ($H$ being a Hubble constant),
in which case
\begin{equation}
\calt_{\Delta,n}=\frac{\Gamma(n+4-\Delta)H^n a^n}{4^n \Gamma(4-\Delta)}\,,\qquad n\ge 0\,.
\eqlabel{tnsolve}
\end{equation}
It is straightforward to verify that the recursive linear ODEs \eqref{feqs} reduce in $\b_2\to 1$
limit to the corresponding equations in \cite{Buchel:2016cbj}. However, for $\b_2\ne 1$ these equations
have more than three singularities on a Riemann sphere; thus, even for $n=0$ case they can only be solved
numerically.

\begin{figure}[t]
\begin{center}
\psfrag{l}{{$\lgb$}}
\psfrag{f}{{$1-\frac{\zeta_3}{{\hat\zeta}_3}$}}
\psfrag{g}{{$1-\frac{\zeta_2}{{\hat\zeta}_2}$}}
  \includegraphics[width=2.4in]{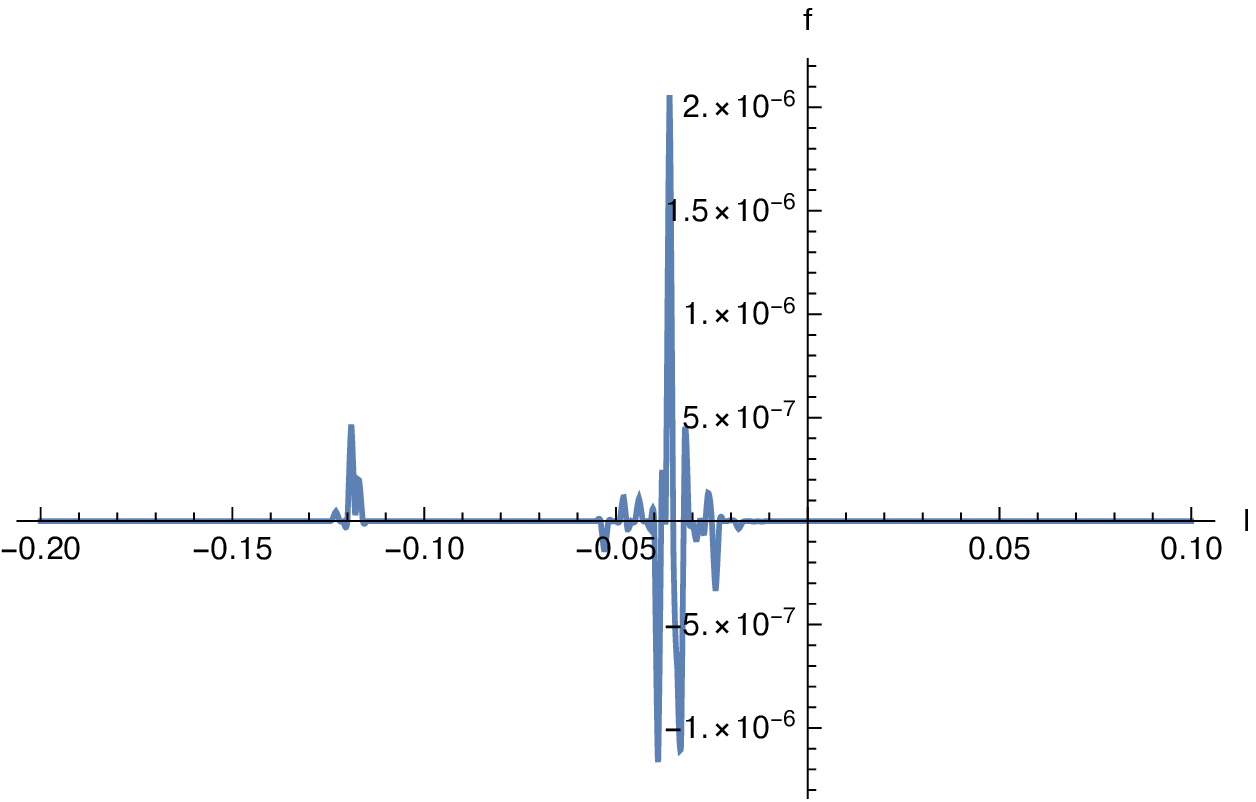}\qquad \qquad
  \includegraphics[width=2.4in]{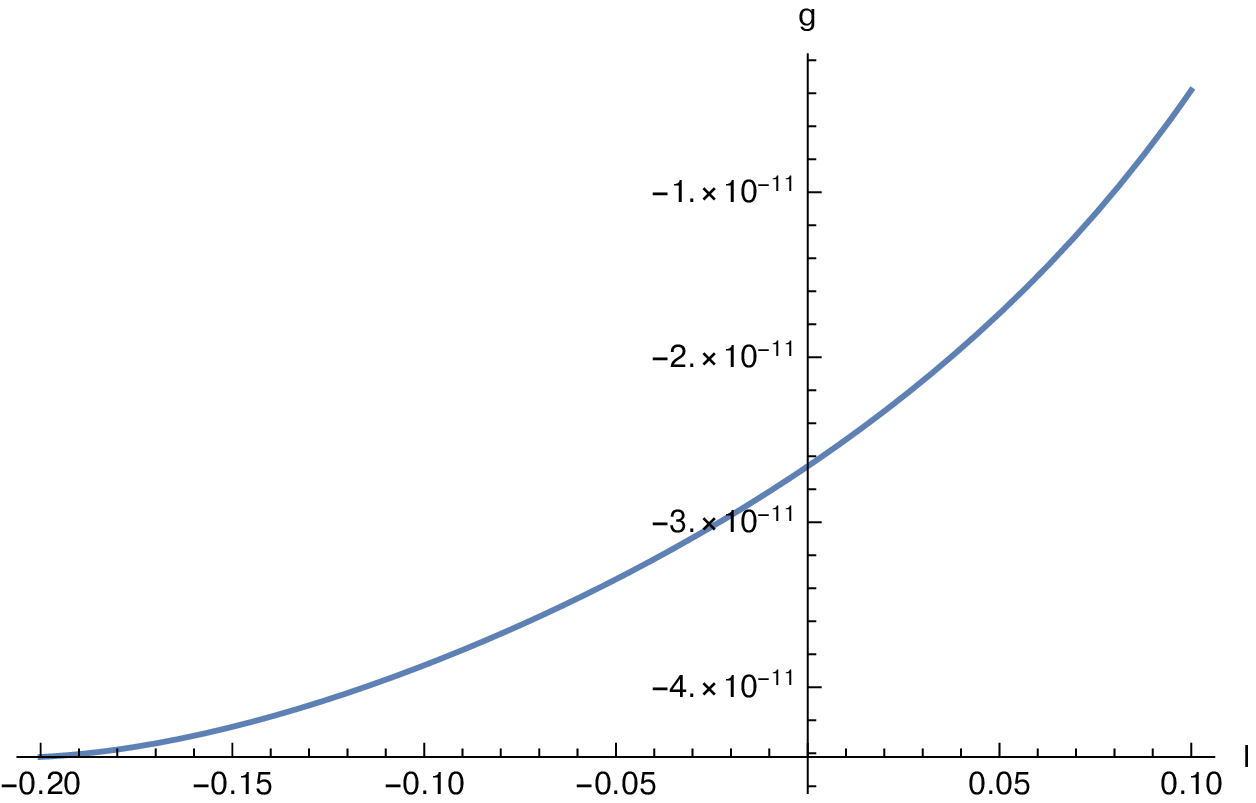}
\end{center}
 \caption{Comparison of the bulk viscosity coefficient $\zeta_\Delta$,
 see \eqref{cszeta}, extracted from the sound waves dispersion relation
and the corresponding coefficient ${\hat\zeta}_\Delta$, see \eqref{zetaetaflrw}, extracted
from the leading hydrodynamic contribution in the entropy production rate
for the FLRW flow. 
}\label{figure7}
\end{figure}

The $n=0$ term in the expansion \eqref{series} represents the leading hydrodynamic response.
Following \eqref{rate}, \eqref{bulkflrw}, \eqref{nonbackreaction}, \eqref{feqs} and \eqref{fbc}
we obtain an elegant expression for the bulk viscosity to the entropy density ratio, to the quadratic order in
the coupling constant $\l_{\Delta-4}$,
\begin{equation}
\frac{\zeta}{s}=\frac{\hdd^2 a^{8-2\Delta}(4-\Delta)^2 \left[F_{\Delta,0}(z=1)\right]^2}{36\pi}=
\frac{(4-\Delta)^2 \left[F_{\Delta,0}(z=1)\right]^2}{\pi^{9-2\Delta}}\ \left(\frac{\l_{4-\Delta}}{T^{4-\Delta}}\right)^2\,,
\eqlabel{zetaflrw}
\end{equation}
correspondingly, using the conformal limit of \eqref{etas},
\begin{equation}
\frac{\zeta}{\eta}=\left(\frac{\l_{4-\Delta}}{T^{4-\Delta}}\right)^2\ \hat{\zeta}_\Delta(\lgb)\,,\qquad
\hat{\zeta}_\Delta=\frac{(4-\Delta)^2}{9\pi^{8-2\Delta}(2\b_2-1)^2}\ \left[F_{\Delta,0}(z=1)\right]^2\,.
\eqlabel{zetaetaflrw}
\end{equation}
Of course, $\hat{\zeta}_\Delta$ should agree precisely with $\zeta_\Delta$ in \eqref{cszeta}.
Fig.~\ref{figure7} demonstrates this agreement. It validates the hydrodynamic computations
in section \ref{hydro}; it also confirms the conjectured identification of the apparent horizon
with the dynamical entropy of the boundary gauge theory in the presence of the bulk GB term.

\begin{figure}[t]
\begin{center}
  \includegraphics[width=2.6in]{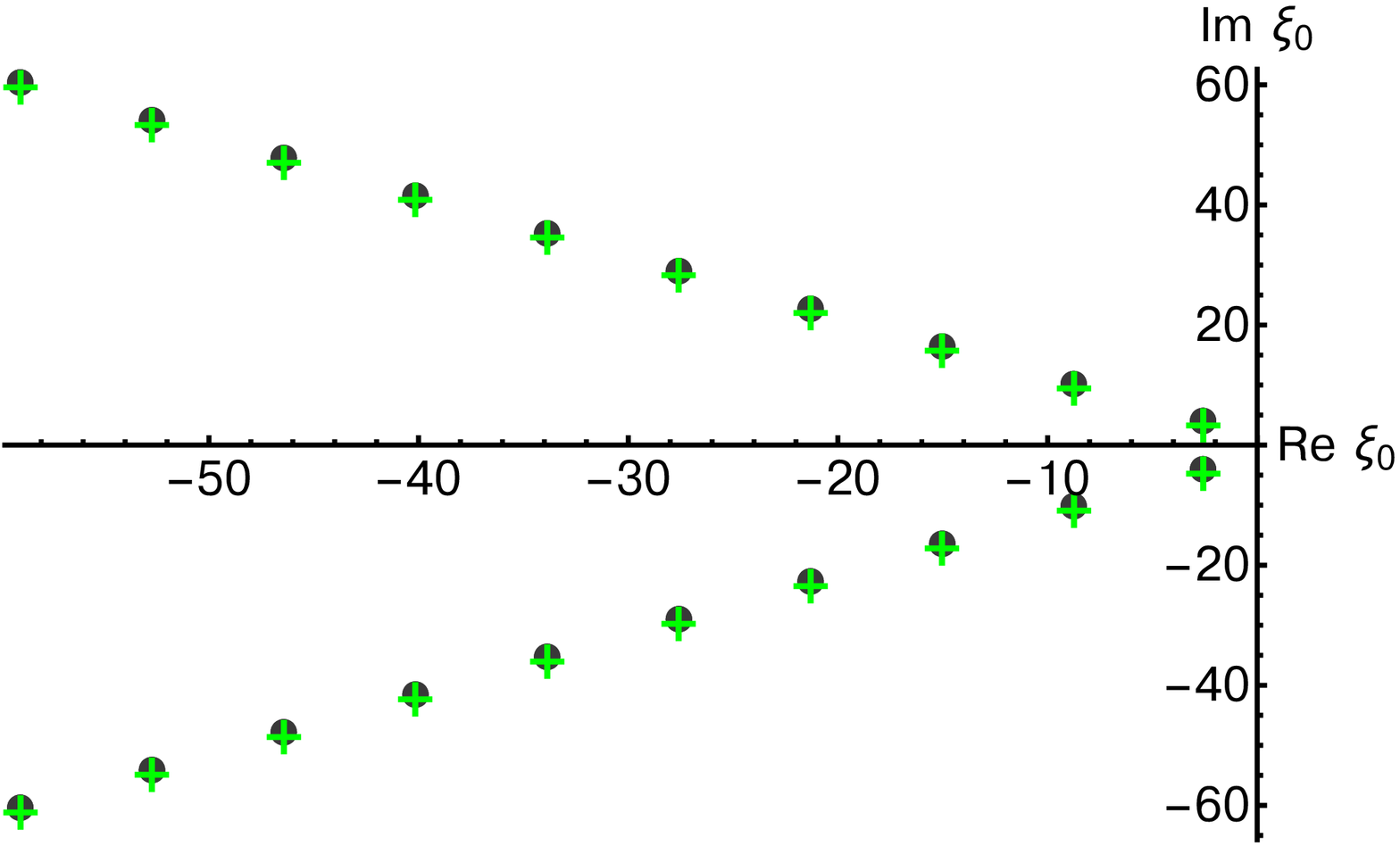}\qquad
  \includegraphics[width=2.6in]{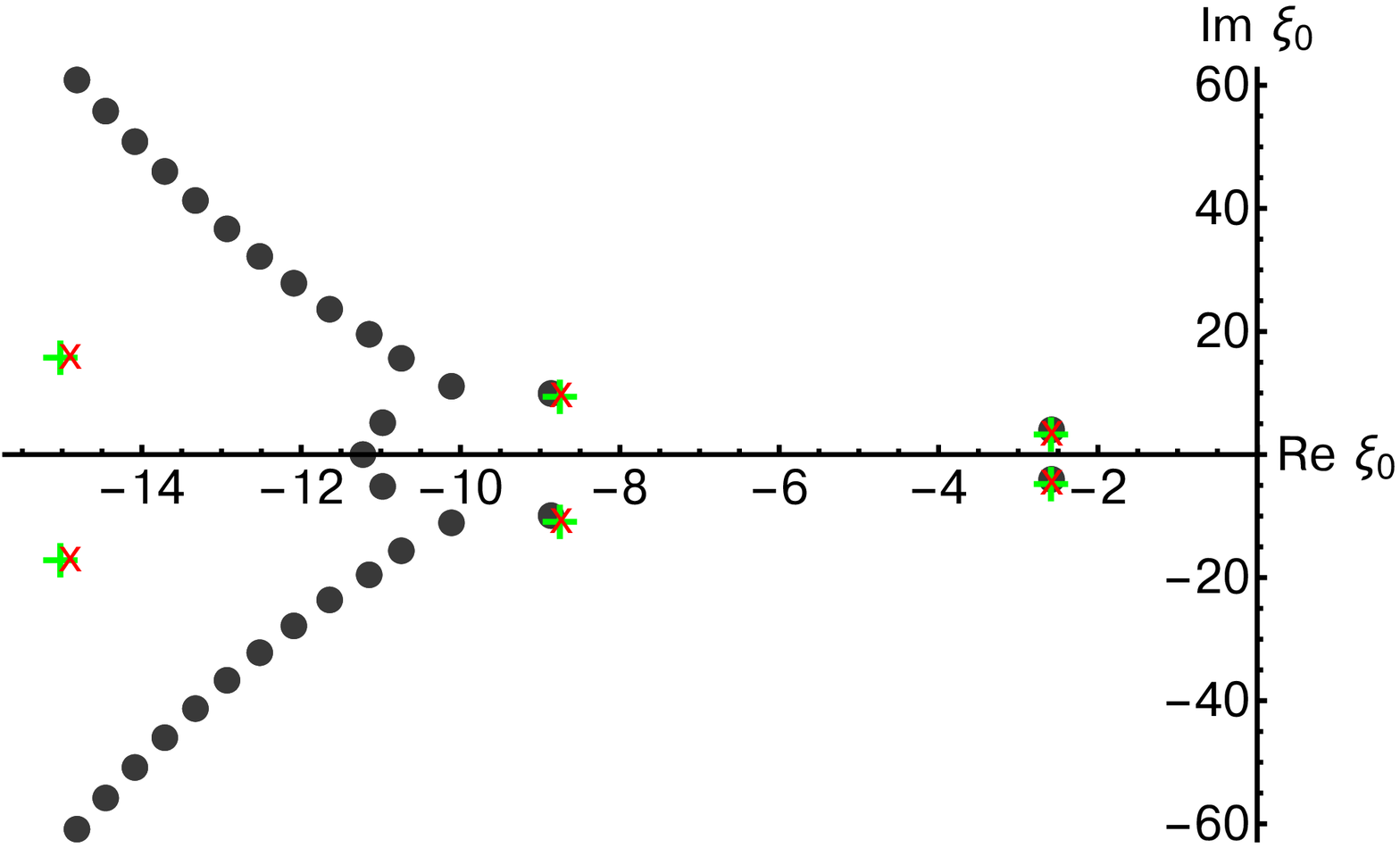}
\end{center}
 \caption{QNMs and leading singularities on the Borel plane for the $\Delta=2$ RG flow with
 $\b_2=1$ (or $\lgb=0$) (left panel)  and  $\b_2=1.001$ (or $\lgb=-0.001001$) (right panel).
We used $n_{max}=300$ terms in the expansion \eqref{series}. See text
 for the legend. 
}\label{figure8}
\end{figure}

\begin{figure}[t]
\begin{center}
  \includegraphics[width=2.6in]{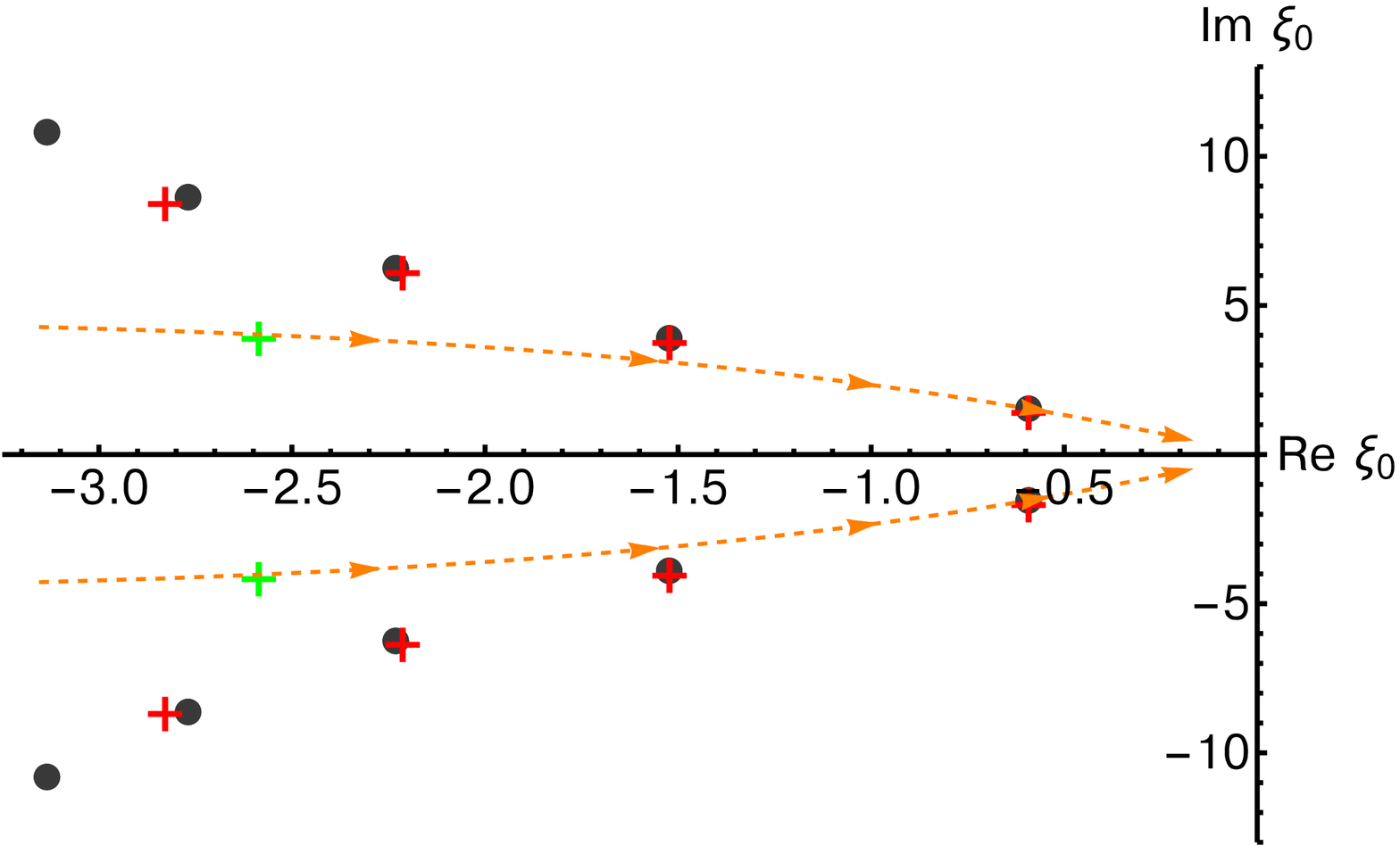}\qquad 
  \includegraphics[width=2.6in]{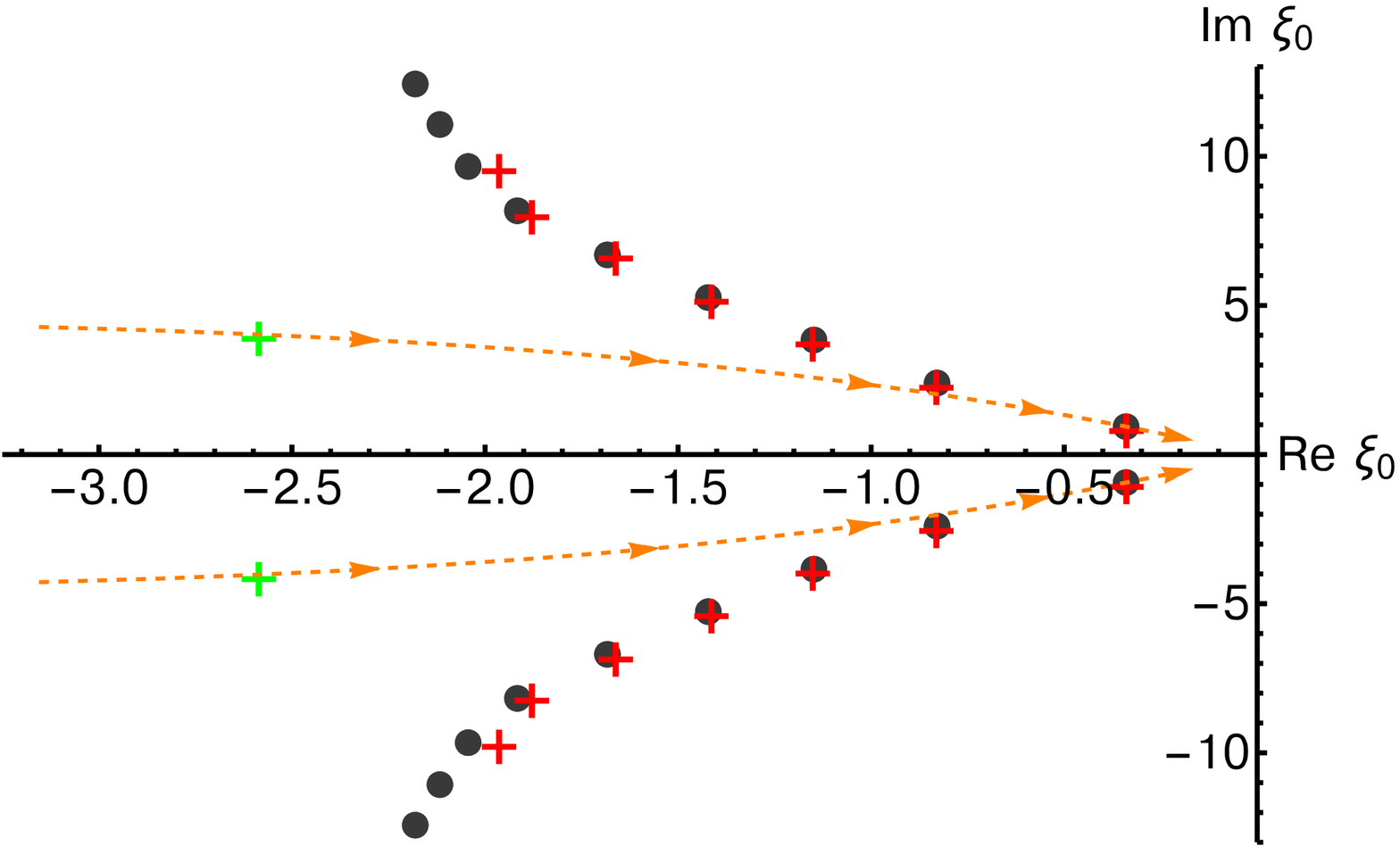}
\end{center}
 \caption{QNMs and leading singularities on the Borel plane for the $\Delta=2$ RG flow with
 $\b_2=3$ (or $\lgb=-6$) (left panel) and $\b_2=5$ (or $\lgb=-20$) (right panel).
 We used $n_{max}=300$ terms in the expansion \eqref{series}. See text
 for the legend. 
}\label{figure9}
\end{figure}

\begin{figure}[t]
\begin{center}
  \includegraphics[width=4.0in]{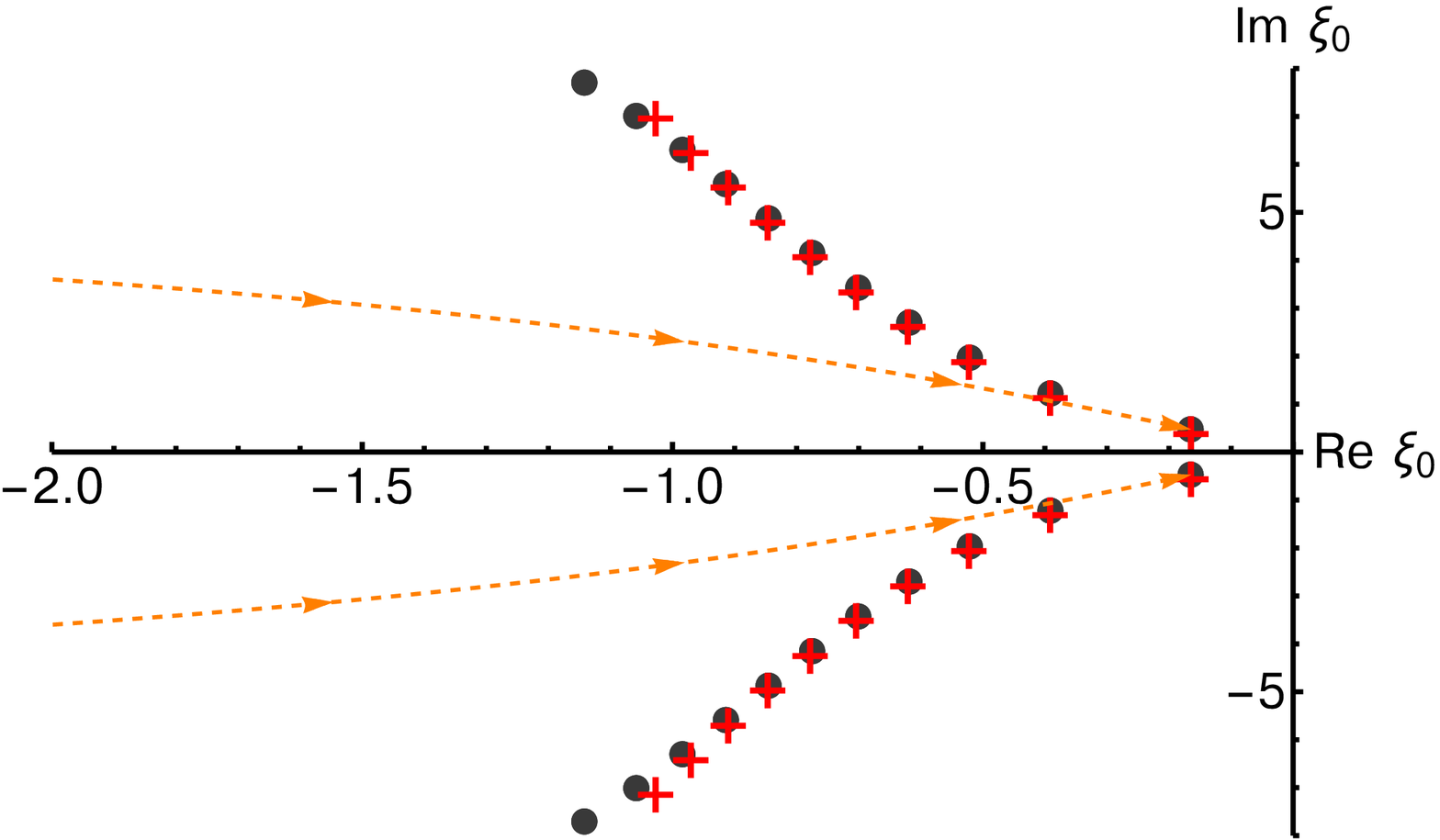}
\end{center}
 \caption{QNMs and leading singularities on the Borel plane for the $\Delta=2$ RG flow with
 $\b_2=10$ (or $\lgb=-90$). We used $n_{max}=300$ terms in the expansion \eqref{series}. See text
 for the legend. 
}\label{figure10}
\end{figure}

We conclude this section commenting on the asymptotic properties of the expansion \eqref{series}.
As argued in \cite{Buchel:2016cbj}, the above expansion is asymptotic at $\b_2=1$, with the
poles in the Pade approximates of their Borel transforms with high accuracy reproducing the
corresponding QNM spectra, \ie the spectra of non-hydrodynamic excitations in the boundary gauge theory plasma.
Here, we have an opportunity to study the interplay of the
convergence of the expansion \eqref{series} and the micro-causality of the model \eqref{uvcausal}.
We focus on $\Delta=2$ RG flows\footnote{There is no qualitative difference for the case of $\Delta=3$.}. 
The results are summarized in figs.~\ref{figure8}-\ref{figure10}:
\begin{itemize}
\item The solid black circles represent the leading singularities $\xi_0$ on the
complex plane closest to the origin for the Borel transform of the expansion
\eqref{series}\footnote{Details of the Borel transform, Pade approximation, {\it etc.}, can be found in
\cite{Buchel:2016cbj}.}. Green crosses correspond to QNM frequencies for $\Delta=2$ and $\b_2=1$
taken from \cite{Nunez:2003eq} and redefined according to $\w_{QNM}(T)=\hw_{QNM} T$ and $\xi_0=-i\hw_{QNM}$.
Red crosses represent QNM frequencies for $\Delta=2$ and $\b_2=\{1.001\,,\, 3\,,\, 5\,,\, 10\}$
(figs.~$\{$\ref{figure8} (right panel)\,,\,
\ref{figure9} (left panel)\,,\, \ref{figure9} (right panel)\,,\,
\ref{figure10}$\}$) correspondingly. Directed orange dashed curves
trace the 'flows' of the lowest QNM mode at $\b_2=1$ to the corresponding value of $\b_2$.
In fig.~\ref{figure9}  the orange flows are extended to $\b_2<1$ ($\lgb>0$)
to illustrate that our computation of the QNMs agrees with results of  \cite{Nunez:2003eq} at $\b_2=1$.
\item The left panel of fig.~\ref{figure8} reproduces the results of \cite{Buchel:2016cbj}
at $\lgb=0$. Computations with small but non-zero $\lgb$ are rather challenging --- the direct
substitution $\b_2=1$ in \eqref{feqs} is {\it singular}, and taking the limit (which as is {\it not singular})
substantially degrades the numerical accuracy. This problem disappears as $\b_2$ deviates substantially from $1$.
Results presented in the right panel of
fig.~\ref{figure8} realize $\lgb=-0.001001$. Because the GB coupling is small, there is almost no flow
for the QNMs: the red and green crosses are on top of each other. Here, we recover only the two lowest
QNMs. We do not believe that 'unmatched' solid circles represent additional QNMs that disappear
in the limit $\b_2\to 1$ (we did not find extra QNMs).
\item Fig.~\ref{figure9} presents the model with $\b_2=3$ (left panel) and $\b_2=5$ (right panel).
Here we reproduce couple more QNMs (red crosses on top of solid circles). Notice that as $\b_2$
increases ($\lgb$ becomes more negative), the singularities on the Borel plane accumulate.
We traced (orange curves) the lowest QNM at $\lgb=0$ (green cross) to the lowest QNM/leading Borel singularity
at corresponding $\lgb$ ($\lgb=-6$ for the left panel and $\lgb=-20$ for the right panel).
We verified that one can trace in a similar fashion higher QNMs as well. Again, we see no indication of
{\it additional} QNMs that are removed from the spectra as $\b_2\to 1$.
\item Fig.~\ref{figure10} presents the model with $\b_2=10$. The general trend observed in fig.~\ref{figure9}
continues: there is even  better agreement between the Borel plane singularities and the QNMs; the singularities
accumulate.
\item We do not present results with $\lgb>0$: they are qualitatively similar to the case of the small negative
GB coupling discussed above. The reason for that is that $0<\lgb<\frac 14$ (correspondingly $1>\b_2>\frac 12$,
\ie $\b_2\sim 1$)
is required for the standard
gauge/gravity dictionary, and at $\lgb=\frac 14$ the GB gravity becomes topological \cite{Chamseddine:1989nu}. 
\end{itemize}
Note: whether or not the model is micro-causal, its hydrodynamic expansion is always asymptotic.

\section{Causality}\label{causality}

Consider a plasma at thermodynamic equilibrium. There will be a spectrum of fluctuations in the plasma, with the dispersion 
relation $\ww=\ww(\kk)$.  The speed
with which a wave-front propagates out from a discontinuity in any
initial data is governed by \cite{fox}
\begin{equation}
\lim_{|\kk|\to\infty}\ \frac{\Re(\ww)}{\kk}= v^{front} \,.
\eqlabel{cs}
\end{equation}
The statement of the microscopic causality of the theory is the statement that for all 
the fluctuations (typically there are multiple branches/channels in the spectrum)
\begin{equation}
v^{front}\le 1\,.
\eqlabel{criteria}
\end{equation}
In the framework of gauge/gravity correspondence, the physical fluctuations in the plasma 
are encoded in the spectrum of the QNMs of the black hole/black brane holographically representing the 
thermal equilibrium state of the latter. For conformal examples of the correspondence with the 
boundary gauge theory having $c-a\ne 0$ micro-causality analysis where performed in 
\cite{Hofman:2008ar,Buchel:2009tt} leading to the constraint \eqref{uvcausal} in GB gravitational models. 
Here we would like to extend the results to non-conformal GB models introduced in section \ref{model}.

Notice that the question of micro-causality is the question of the deep UV properties of the theory, 
thus one expects that breaking the scale invariance with a relevant or marginal deformation, \ie with a  dimension 
$\Delta\le 4$ operator, should not affect the result \eqref{uvcausal}. Causality should not 
depend on the state of the theory\footnote{We explicitly verified this statement in our models.}, 
for example, the temperature compare to the coupling 
strength  $\l_{4-\Delta}$. 
However, in principle,
\begin{itemize} 
\item If several relevant couplings are present, causality can be affected by the dimensionless 
ratio of these couplings\footnote{We can not probe this in our GB models, as we have a single relevant deformation 
which is treated in the probe approximation.}.
 \item Additionally, recall \cite{Buchel:2009tt} that different channels of the 
fluctuations in plasma affect causality differently: the scalar channel 
of the bulk graviton fluctuations constraints  
\begin{equation}
\lgb\le \lgb^{scalar}= \frac{9}{100}\,,
\eqlabel{lscalar}
\end{equation}
 while the shear and the sound channels constraint correspondingly:
\begin{equation}
\lgb\ge \lgb^{shear}=-\frac 34\,,\qquad  \lgb \ge \lgb^{sound}=-\frac{7}{36}\,.
\eqlabel{critlgb}
\end{equation}
It is only the union of all the constraints that determines \eqref{uvcausal}. If the theory is non-conformal, obviously, 
there is a spectrum of operators present at its UV fixed point, which coupling constants can be adjusted. Existence of 
these operators introduces additional fluctuation channels (additional branches of the QNMs) which can further 
constraint the microscopic causality of the model. 
\end{itemize} 

In the section we investigate the second of the possibilities mentioned above. To this end,  
consider the branch of the QNMs of the 'conformal' black brane geometry, \ie \eqref{metric} with \eqref{cft}
for the metric warp factors, associated with the fluctuations of the bulk scalar field, dual to a dimension 
$\Delta\le 4$ operator. Following \cite{Buchel:2009tt}, this quasinormal mode equation can be rewritten in the form 
of the Schr\"odinger equation:
\begin{equation}
\begin{split}
&-\hbar^2\ \del_y^2\, \psi_{[\Delta]} +U_{[\Delta]}\
\psi_{[\Delta]}
=\a^2\ \psi_{[\Delta]}\,,\qquad \hbar\equiv \frac {1}{\kk}\,,\qquad \a=\frac\ww\kk\,,\\
&\qquad{\rm where}\ \ \ U_{[\Delta]}=U^0_{[\Delta]}+\hbar^2\
U^1_{[\Delta]}\,.
\end{split}
\eqlabel{sshear}
\end{equation}
The first part of the effective potential has the simple form when
expressed in terms of $x$,
 \begin{equation}
U^0_{[\Delta]}(x)=\frac{\sqrt{(2\b_2-1)^2-4 \b_2(\b_2-1) x^2}-1}{2(\b_2-1)}\,.
\eqlabel{u0sh}
\end{equation}
while the expression for $U^1_{[\Delta]}$ is too long to be presented
here, but we note that the latter is a function only of $x$, $\b_2$,
$\Delta$ and $\alpha$. What is important is that in the limit $\kk\to \infty$
(or $\hbar\to0$), everywhere except in the tiny region $y\gtrsim
-\frac{1}{\kk}$ the dominant contribution to $U_{[\Delta]}$ comes from
$U^0_{[\Delta]}$. Thus in this limit we simply replace
\begin{equation}
\hbar^2\, U^1_{[\Delta]}=\begin{cases} 0 & \text{$y<0$\,,}
\\
+\infty &\text{$y\ge0$\,.}
\end{cases}
\eqlabel{u1shear}
\end{equation}
Causality is violated if the effective Schr\"odinger problem has a bound state with $\a^2>1$. It is easy to see
that such a bound state
does not exist for any value of $\b_2$ since $U^0_{[\Delta]}$ is a monotonically decreasing function of $x$ from $1$ to $0$. 
We conclude that the spectrum of operators of a GB CFT (besides the stress-energy tensor) does not further constraint its causal properties 
beyond \eqref{uvcausal}.

\section{Conclusion}\label{summary}

In this work we summarized some (near-)equilibrium 
properties of the (phenomenological) holographic RG flows with a dual four-dimensional gauge theory interpretation. 
The UV fixed point of the theory has different central charges, \ie  $c-a\ne 0$, and the flow is 
triggered by the relevant operator $\calo_\Delta$ with dimension $\Delta=\{2,3\}$. We considered RG flows close to the UV fixed, \ie 
the mass scale associated with the coupling constant of the conformal symmetry breaking deformation, $\l_{4-\Delta}$, is much smaller 
that the local temperature of the boundary gauge theory plasma, see \eqref{close}. 
We worked to leading nontrivial order in the (explicit) conformal
symmetry breaking parameter, but for arbitrary finite values of $c-a$. The simple gravitational model capturing the physics is that 
of the five-dimensional GB gravity with a minimally coupled bulk scalar field of the appropriate mass, see \eqref{action}.

To summarize:
\nxt We presented holographic renormalization of the model, sufficient to compute the one- and two-point thermal correlation functions 
of the stress-energy tensor and $\calo_\Delta$. 
\nxt We computed equation of the state, the transport properties 
(the speed of sound waves, the shear and bulk viscosities), and studied the large-order hydrodynamic gradient expansion in our GB plasma.  
We discussed the micro-causality of the model. 
\nxt Particular attention was devoted towards consistencies of the computations: the holographic renormalization was checked 
testing the first law of thermodynamics (see fig.~\ref{figure2}); 
the speed of sound waves was computed from the equation of state \eqref{defcs} and compared with leading-order term 
in the sound-channel QNMs dispersion relation \eqref{sounddisp} (see fig.~\ref{figure4});  the bulk viscosity
was extracting from the sound waves dispersion relation \eqref{sounddisp} and compared with the bulk viscosity
obtained from the entropy growth rate for the homogeneous and isotropic expansion of the plasma \eqref{bulkflrw}
(see fig.~\ref{figure7}) --- notice that because the attenuation of sound waves depends on the shear viscosity of the plasma as
well, we are indirectly testing here the consistency of the shear viscosity computations from the sound waves and from the 
viscosity Kubo formula \eqref{toff}. We verified that our transport coefficients (and the spectrum of the
non-hydrodynamic modes in plasma) at $c=a$ agrees with the results (whenever available) in the literature. 
\nxt We presented a simple and compact formula, see \eqref{zetaetaflrw}, for the bulk viscosity
from the entropy growth rate in GB model, reminiscent to the Eling-Oz formula \cite{Eling:2011ms,Buchel:2011yv}.
\nxt We argued that non-conformal deformations of a holographic CFT with $c-a\ne 0$ do not effect the causal properties of the
theory --- allowed range of the GB coupling constant (or the difference of the central charged at the UV fixed point) is still
given by \eqref{uvcausal}.
\nxt We showed that the bulk viscosity bound introduced in \cite{Buchel:2007mf} is violated for sufficiently
large $a-c>0$. This should be contrasted with the shear viscosity bound \cite{Kovtun:2004de} which is violated
for arbitrary small $c-a>0$. 
\nxt Conformal field theories have vanishing $\dd_{c_s}\equiv\frac 13-c_s^2$ and $\dd_\zeta
\equiv \frac{\zeta}{\eta}$. In non-conformal RG flows both $\dd_{c_s}$ and $\dd_{\zeta}$ do not vanish:  while the former
has a rather mild dependence on $\lgb$ in the causal windows ( $\dd_{c_s}$ varies by $\sim 15\%$ for $\Delta=2$ deformation,
and by $\sim 8\%$ for $\Delta=3$ deformation ), the variation of the latter is more substantial  (
$\dd_{\zeta}$ varies by $\sim 50\%$ for $\Delta=2$ deformation,
and by $\sim 60\%$ for $\Delta=3$ deformation). Shear viscosity does not vanish in the conformal limit and varies
by $\sim 80\%$ in the causal window. 
\nxt We showed that the hydrodynamic expansion in non-conformal GB models is an asymptotic series,
whether or no the model is microscopically causal. As in \cite{Buchel:2016cbj}, for $\lgb\ne 0$ (or $c-a\ne 0$)
the leading singularities on the Borel plane for non-conformal RG flows agree with the corresponding QNMs --- the agreement
improves as $(-\lgb)$ becomes larger. We observe accumulation of the singularities  close to the origin for
large $a-c>0$. Our analysis support the physical picture advocated in the original work \cite{Heller:2013fn}
that the asymptotic properties of the hydrodynamic gradient expansion are controlled by the non-hydrodynamic modes,
with the lowest lying modes being the most important. Thus our
results are not surprising: low-lying states in the spectrum of non-hydrodynamic excitations in plasma do not
probe the micro-causality of the model. This picture has further nice confirmation in the recent work 
\cite{Rozali:2017bll}. Here, the full hydrodynamic gradient expansion truncates at the second order (being obviously
a convergent series) because the limit of the large number of spatial dimensions decouples (removes from the spectrum)
the non-hydrodynamics plasma excitations (non-hydrodynamic QNMs in the dual gravitational description).

\section*{Acknowledgments}
Research at Perimeter
Institute is supported by the Government of Canada through Industry
Canada and by the Province of Ontario through the Ministry of
Research \& Innovation. This work was further supported by
NSERC through the Discovery Grants program.

%\appendix
%\section{Apparent horizon area growth theorem}\label{th1}

\end{document}